\newcommand{\CLASSINPUTbottomtextmargin}{1.8cm}
\documentclass[journal]{IEEEtran}
\usepackage{stackengine}
\usepackage{amssymb}
\usepackage{amsmath}
\usepackage{dsfont}
\newcommand \ubar[1]{\stackunder[1.1pt]{$#1$}{\rule{.8ex}{.075ex}}}
\usepackage{amssymb}
\usepackage{cite}
\usepackage{graphicx}
\usepackage{bm}
\usepackage{multirow}
\usepackage{booktabs}
\usepackage{color}
\usepackage{upgreek}
\usepackage[bookmarks={false}]{hyperref}
\usepackage[hyphenbreaks]{breakurl}
\usepackage{balance}
\usepackage{verbatim}
\usepackage{relsize}
\usepackage{nomencl}
\usepackage{ifthen}
\usepackage{soul}

\usepackage{rotating}
\usepackage{accents}
\usepackage{setspace}
\usepackage{bbold}
\usepackage[compact]{titlesec}
\usepackage{tikz}
\usepackage{amssymb}
\usepackage{cuted}
\usepackage{float}
\usepackage{stfloats}
\usepackage{nicematrix}
\newcommand{\wye}{\mathbin{\tikz[x=0.6ex,y=0.6ex]{\draw[line width=.1ex] (0,0)--(-30:1)--++(30:1) (-30:1)--++(0,-1) ;}}}

\begin{document}
\title{Generalized Analytical Estimation of Sensitivity Matrices in Unbalanced Distribution Networks}

\author{Salish~Maharjan,~\IEEEmembership{Member,~IEEE,} Rui Cheng,~\IEEEmembership{Student~Member,~IEEE,}and
        Zhaoyu~Wang,~\IEEEmembership{Senior~Member,~IEEE}
        \vspace{-1cm}
}

\markboth{Journal of \LaTeX\ Class Files,~Vol.~14, No.~8, August~2015}%
{Shell \MakeLowercase{\textit{et al.}}: Bare Demo of IEEEtran.cls for IEEE Journals}

\maketitle

\begin{abstract}
Fast and accurate estimation of sensitivity matrices is significant for the enhancement of distribution system modeling and automation. Analytical estimations have mainly focused on voltage magnitude sensitivity to active/reactive power injections for unbalance networks with Wye-connected loads and neglecting DERs’ smart inverter functionality.  Hence, this paper enhances the scope of analytical estimation of sensitivity matrices for unbalanced networks with 1-$\phi$, 2-$\phi$, and 3-$\phi$ Delta/Wye-connected loads, DERs with smart inverter functionality, and substation/line step-voltage regulators (SVR).  A composite bus model comprising of DER, Delta- and Wye-connected load is proposed to represent a generic distribution bus, which can be simplified to load, PV, or voltage-controlled bus as required.  The proposed matrix-based analytical method consolidates voltage magnitude and angle sensitivity to active/reactive power injection and tap-position of all SVRs into a single algorithm. Extensive case studies on IEEE networks show the accuracy and wide scope of the proposed algorithm compared to the existing benchmark method.
\end{abstract}


\IEEEpeerreviewmaketitle

\section{Introduction}
Broadly, sensitivity coefficients are defined by the first-order partial derivative of any state variables to the input variable. In particular to the distribution network, sensitivity coefficients generally refer to the partial derivative of nodes' voltage magnitude ($\mathbf{E}$) and angle ($\boldsymbol{\theta}$) to active/reactive nodal power injections ($\mathbf{P}/\mathbf{Q}$) and tap-position ($\boldsymbol{\gamma}$) of voltage regulators, i.e., $\frac{\partial \mathbf{E}}{\partial \mathbf{P}}$, $\frac{\partial \mathbf{E}}{\partial \mathbf{Q}}$, $\frac{\partial \mathbf{E}}{\partial \boldsymbol{\gamma}}$,  $\frac{\partial \boldsymbol{\theta}}{\partial \mathbf{P}}$, $\frac{\partial \boldsymbol{\theta}}{\partial \mathbf{Q}}$, and $\frac{\partial \boldsymbol{\theta}}{\partial \boldsymbol{\gamma}}$. Sensitivities to other network's state variables, such as line current and loss, are computed using voltage magnitude and angle sensitivities \cite{Zhou2008SimplifiedNetworks}. Estimation of voltage and angle sensitivities are generally provided as a built-in function in transmission network simulation tools (such as MATPOWER and DigSILENT) which typically employ the Jacobian method \cite{Zimmerman2019ACNotation}. In contrast, currently available distribution network simulation tools, such as Open DSS, PandaPower, and DigSILENT, do not have built-in functions to estimate voltage and its angle sensitivities for the unbalanced system. It is mainly due to the complexity of distribution network modeling and solving in the presence of multi-phased lines, loads, and distributed energy resources (DERs) and their various configurations (e.g., variants of Delta and Wye connections).
\par
Network sensitivities have been popularly used to achieve closed-loop control of distribution networks for achieving voltage control \cite{Maharjan2021RobustNetworks}, optimal economic operation \cite{Jiang2019StochasticRenewables},  catering ancillary services \cite{Gupta2020Grid-awareValidation}, and safely re-closing breakers \cite{Picallo2022AdaptiveEstimation}. With the increasing penetration of renewable DERs and electric vehicles in distribution systems, online feedback optimization is regularly used to respond quickly to network changes. Majorly online feedback optimization is dependent on sensitivities \cite{Picallo2022AdaptiveEstimation,Haberle2021Non-ConvexConstraints}. Hence, fast and accurate estimation of sensitivities is of significant importance for the enhancement of distribution network automation.

\par

The methods to estimate voltage sensitivities can be broadly classified into two categories, viz., (a) data-driven and (b) model-based. Data-driven methods are typically neural networks trained to predict the sensitivities at various operating conditions \cite{Shi2022Data-DrivenControl}.
However, Data-driven methods always require high-quality and large datasets, and they are difficult to reveal physical laws. Instead, analytical methods are physics-based, which do not depend on high-quality and large datasets. Analytical methods can be further classified into two categories based on their application to only radial networks \cite{Conti2010VoltageModels, Ouali2018SensitivitySystems,Kumar2005VoltageApproach,Hong2014AnSystems,BakhshidehZad2015OptimalAlgorithm,Zad2018AImpact} and to both radial and meshed distribution networks \cite{Zhou2008SimplifiedNetworks,Christakou2013EfficientNetworks,Maharjan2020,Fahmy2021AnalyticalInjections,KumarR2022NeumannSystem}.


\par
The study in \cite{Conti2010VoltageModels, Ouali2018SensitivitySystems} proposes a simplified approach to compute voltage sensitivity coefficients in radial distribution networks for constant current loads/sources and is further simplified by neglecting phase differences among buses. However, network control and operation consider a constant power model of loads/sources, which limits the application of these methods. Considering constant power models, the voltage sensitivity coefficients for radial systems are analytically formulated in \cite{Kumar2005VoltageApproach,Hong2014AnSystems,BakhshidehZad2015OptimalAlgorithm,Zad2018AImpact}. The estimated sensitivities in \cite{Kumar2005VoltageApproach,Hong2014AnSystems,BakhshidehZad2015OptimalAlgorithm}  are exact for the radial lossless networks and are generalized in \cite{Zad2018AImpact} considering the line losses.
\par
The analytical methods applicable to both radial or meshed distribution networks typically employ Z-matrix \cite{Zhou2008SimplifiedNetworks,Maharjan2020}, Y-matrix\cite{Christakou2013EfficientNetworks,Fahmy2021AnalyticalInjections},  or both Y- and Z-matrix \cite{KumarR2022NeumannSystem} to express the relationship between  power injections and node/bus voltages for sensitivity estimation. Formative work on sensitivity estimation of distribution networks is conducted in \cite{Zhou2008SimplifiedNetworks}, where the first-order partial derivatives of bus voltage with respect to active/reactive power injections are estimated by solving linear sets of equations pertaining to nodal power injections. This study is further enhanced in \cite{Maharjan2020} by integrating voltage sensitivity to the tap-position of a substation transformer in distribution networks and by demonstrating its applicability in meshed networks. However, the application of both works \cite{Zhou2008SimplifiedNetworks,Maharjan2020} are limited to balanced distribution networks only. An influential work on voltage sensitivity estimation in unbalanced distribution networks is studied in \cite{Christakou2013EfficientNetworks} considering multiple slack and load buses. The work is further generalized in \cite{Fahmy2021AnalyticalInjections} by considering PV buses too. A Neumann series is applied in \cite{KumarR2022NeumannSystem} to simplify voltage sensitivity estimation to active/reactive power injections. The works \cite{Christakou2013EfficientNetworks,Fahmy2021AnalyticalInjections,KumarR2022NeumannSystem} only consider  Wye-connected loads. However, there are also Delta-connected loads that could be either 1-$\phi$, 2-$\phi$, or 3-$\phi$ in reality. Furthermore, the DERs' smart inverter functionalities, such as volt-var control, are also not considered in \cite{Christakou2013EfficientNetworks,Fahmy2021AnalyticalInjections,KumarR2022NeumannSystem}. The smart inverter functionalities directly impact the network voltage and are popularly used as recommended by IEEE 1547-2018 \cite{IEEEStandardAssociation2018IEEEInterfaces}. Hence at large DER penetration, it is significant to consider DERs' smart inverter functionality while estimating the sensitivity matrices. There are multiple 1-$\phi$ substation/line step-voltage regulators (SVR) deployed mainly for voltage control, and the voltage sensitivities to the tap-positon of such SVRs are not yet studied in the past literature. 

\par
Hence, this paper enhances the analytical estimation of voltage sensitivity matrices in unbalanced distribution networks considering DERs' smart inverter functionalities, multi-phased Delta/Wye connected loads, and substation/line SVRs. Furthermore, the proposed matrix-based method consolidates voltage magnitude and angle sensitivity to active/reactive power injection and to tap-position of substation/line SVRs in a single algorithm. The wide applicability of the proposed algorithm is achieved by modeling each bus as a composite bus comprising DER, Delta-, and Wye-connected loads. The composite bus represents the reality of distribution buses as there is no definite load and generator bus in the distribution system. The composite bus can be easily simplified to a generator (PV bus), load, or voltage-controlled bus as required.  The proposed algorithm is tested in various IEEE networks, and the performance is evaluated by mean absolute percentage error and mean computation time. 
\par
The contributions of the paper are listed as follows:
\begin{itemize}
    \item Formulate \textit{a generalized analytical method}  for voltage magnitude and angle sensitivity matrices with respect to active/reactive power injections and tap-position of SVRs.
    \item This proposed analytical method extends sensitivity matrices to more realistic and comprehensive distribution networks, considering not only SVRs but also multi-phase Delta- and Wye-connected loads.
    \item This proposed analytical method takes into account DERs' smart inverter functionalities, greatly improving  its generalization ability and flexibility.
\end{itemize}

\section{Analytical derivation of sensitivity matrices}
\subsection{Modeling unbalanced distribution network in matrix form}
For a general three-phase distribution network, the injected node currents and node voltages are linked by its admittance matrix as\footnote{In this paper, every phasor, its conjugate and magnitude are denoted with a bar above (e.g., $\bar{X}$), below (e.g., $\ubar{X}$) and without any bars (e.g., $X$), respectively. Additionally, normal matrix multiplication is denoted by $\cdot$.}:
\begin{align}
    \mathbf{\bar{I}}=\mathbf{\bar{Y}}\cdot\mathbf{\bar{E}}. \label{eq_IeqYE}
\end{align}
Here, $\mathbf{\bar{I}}=[\bar{I}_{a}^1,\bar{I}_{b}^1,\bar{I}_{c}^1,\dots,\bar{I}_{a}^n,\bar{I}_{b}^n,\bar{I}_{c}^n]^T$, and $\mathbf{\bar{E}}=$ $[\bar{E}_{a}^1,\bar{E}_{b}^1,\bar{E}_{c}^1,\dots,\bar{E}_{a}^n, \bar{E}_b^n, \bar{E}_{c}^n]^T$. Here, the super-scripts $\{1,2,\dots,n\}$ denote the bus, whereas the sub-scripts $\{a,b,c\}$ represent nodes/phases associated with the corresponding bus. The system admittance matrix $\mathbf{\bar{Y}}$ is formed by clustering the primitive admittance matrix of each network element such as lines, switches, capacitor banks, transformers, and regulators (e.g.,\cite{Kersting2018DistributionAnalysis}), and  has the structure as follows:
\begin{align}
    \mathbf{\bar{Y}}=
\begin{bmatrix}
\bar{y}_{aa}^{11}\:\:\bar{y}_{ab}^{11}\:\:\bar{y}_{ac}^{11} & \cdots & \bar{y}_{aa}^{1n}\:\:\bar{y}_{ab}^{1n}\:\:\bar{y}_{ac}^{1n} \\
\bar{y}_{ba}^{11}\:\:\bar{y}_{bb}^{11}\:\:\bar{y}_{bc}^{11} & \cdots & \bar{y}_{ba}^{1n}\:\:\bar{y}_{bb}^{1n}\:\:\bar{y}_{bc}^{1n} \\
\bar{y}_{ca}^{11}\:\:\bar{y}_{cb}^{11}\:\:\bar{y}_{cc}^{11} & \cdots & \bar{y}_{ca}^{1n}\:\:\bar{y}_{cb}^{1n}\:\:\bar{y}_{cc}^{1n} \\
\vdots\quad\:\:\: \vdots \quad\:\:\: \vdots & \ddots &\vdots\quad\:\:\: \vdots \quad\:\:\: \vdots \\
\bar{y}_{aa}^{n1}\:\:\bar{y}_{ab}^{n1}\:\:\bar{y}_{ac}^{n1} & \cdots & \bar{y}_{aa}^{nn}\:\:\bar{y}_{ab}^{nn}\:\:\bar{y}_{ac}^{nn}\\
\bar{y}_{ba}^{n1}\:\:\bar{y}_{bb}^{n1}\:\:\bar{y}_{bc}^{n1} & \cdots & \bar{y}_{ba}^{nn}\:\:\bar{y}_{bb}^{nn}\:\:\bar{y}_{bc}^{nn}\\
\bar{y}_{ca}^{n1}\:\:\bar{y}_{cb}^{n1}\:\:\bar{y}_{cc}^{n1} & \cdots & \bar{y}_{ca}^{nn}\:\:\bar{y}_{cb}^{nn}\:\:\bar{y}_{cc}^{nn}
\end{bmatrix}
\end{align}
\par
The distribution network comprises a few three-phase and single-phase tap-changing transformers (also referred to as step-voltage regulators) at the substation or along the line, primarily designed for voltage regulation. As a result, its $\mathbf{\bar{Y}}$ has to be recomputed whenever the tap is shifted in those transformers. To avoid the entire recomputation of the admittance matrix, we formulate it as:
\begin{align}
    \mathbf{\bar{Y}} = \mathbf{\bar{Y}^o} + \mathbf{\bm{\delta}\bar{Y}^r},\label{eq_add_mat}
\end{align}
where $\mathbf{\bar{Y}^o}$ is the admittance matrix when the taps are at the nominal position and remain constant unless the distribution network is reconfigured. $\mathbf{\bm{\delta}{\bar{Y}}^r}$ is an incremental change in the admittance matrix, which accounts for the change in admittance due to shifting in the tap-position of regulators. $\mathbf{\bm{\delta}\bar{Y}^r}$ is highly sparse than $\mathbf{\bar{Y}^o}$, and can be computed based on location and type of regulator as\footnote{The nodes are represented by a tuple (bus, phase), where bus refers to bus name or number and phase is either $a$, $b$ or $c$.}:
\begin{align}
&\text{For all } \left((i,p),(j,k)\right)\in\mathcal{R}\nonumber\\
    &\quad \quad\mathbf{\bm{\delta}\bar{Y}^r}\left((i,p),(j,k)\right) = \delta\bar{Y}^{r}\left((i,p),(j,k)\right)\nonumber\\
&\text{For all } \left((i,p),(j,k)\right)\notin\mathcal{R}\nonumber\\
&\quad\quad\mathbf{\bm{\delta}\bar{Y}^r}\left((i,p),(j,k)\right) = 0\nonumber
\end{align}
Here, $(i,p)$ and $(j,k)$ represent the `from' and `to' nodes of a voltage regulator, where $i$ and $j$ are bus indices, and $p$ and $k$ are phase indices. $\mathcal{R}$ is a set containing the 'from' and 'to' nodes of all voltage regulators in the distribution network. $\delta \bar{Y}^r$ is an incremental admittance matrix of each voltage regulator with reference to its admittance matrix at the nominal tap position. $\bm{\delta}\bar{Y}^{r}$ for different types of regulator are shown in Appendix \ref{app_del_Y_1_phase}, \ref{app_del_Y_3_phase}. 
\par
\begin{figure}[t]
    \centering
    \includegraphics[width=0.65\linewidth]{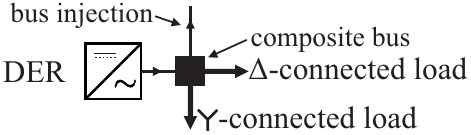}
       \caption{Composite bus model} \label{fig_composite_bus}
\end{figure}
The distribution network comprises one or more interconnection with the transmission grid, referred to as slack buses, and their set of nodes are represented by $\mathcal{S}$. Unlike transmission networks, distribution networks do not have a distinct generator or load bus, rather both generators (referred to as DERs) and load co-exist on the same bus. In addition, the DERs and loads could be either three-phase, two-phase, or single-phase in nature. Again, the loads could be either Wye or Delta connected, whereas DERs are generally Wye connected \cite{Peterson2019AnSolutions}. Consequently, the nodal current injection on distribution networks is more heterogeneous than in transmission systems. Without loss of generality, the distribution network bus injection is modeled by three-phase Delta and Wye-connected loads, and Wye-connected DER, as shown in Fig. \ref{fig_composite_bus}. This composite injection model can be easily simplified to any 1-$\phi$ or 2-$\phi$ or 3-$\phi$ connection of loads and DERs. Furthermore, the composite bus can also be simplified to model a generator bus (or PV bus) or a load bus, or a voltage control bus. The set of nodes of composite buses is represented by $\mathcal{C}$.
\par
For the sake of generic modeling, all the buses of the distribution network are considered to be composite buses. However, the slack buses will be considered later while solving the network power flow by converting their composite model to a slack bus. With this proposition, the current injection vector of the distribution network can be written as:
\begin{align}
    -\mathbf{\bar{I}}_{L,{\wye}}-\mathbf{\bar{I}}_{L,\Delta\text{-}\wye}+\mathbf{\bar{I}}_{G} =(\mathbf{\bar{Y}}^{o}+\mathbf{\bm{\delta}\bar{Y}^r})\cdot\mathbf{\bar{E}}.\label{eqn_Iinj}
\end{align}
Here $\mathbf{\bar{I}}_{L,{\wye}}$ and $\mathbf{\bar{I}}_{G}$ are vectors of current injection from Wye-connected loads and DERs, respectively. Meanwhile, $\mathbf{\bar{I}}_{L,\Delta\text{-}\wye}$ \textcolor{red}{is} a vector of current injection from Wye-transformed Delta-connected loads. Note that (\ref{eqn_Iinj}) holds true only for nodal current injections. Star-connected loads are inherently nodal injections, however, delta-connected loads inject the current across the phases. Hence, all delta-connected loads are required to be converted to equivalent star-connection ($\mathbf{\bar{I}}_{L,\Delta\text{-}\wye}$) before equating them in (\ref{eqn_Iinj}).
The current injection form of (\ref{eqn_Iinj}) can be expressed in terms of complex power injection as \footnote{The Hadamard product is denoted by $\odot$ throughout the paper.}:
\begin{align}
    -\ubar{\mathbf{S}}_{L,{\wye}}-\ubar{\mathbf{S}}_{L,\Delta\text{-}\wye} +  \ubar{\mathbf{S}}_G=\ubar{\mathbf{E}}\odot(\mathbf{\bar{Y}}^{o}+\mathbf{\bm{\delta}\bar{Y}^r})\cdot\mathbf{\bar{E}} \label{eqn_Sinj}.
\end{align}
Here $\ubar{\mathbf{S}}_{L,{\wye}}$ and $\ubar{\mathbf{S}}_G$ are complex conjugate of power injection vector of Wye-connected loads and DERs, respectively. $\ubar{\mathbf{S}}_{L,\Delta\text{-}\wye}$ is a Wye-transformed load vector which is obtained by transforming the vector of Delta load to Wye connection.

\subsubsection{Transforming vector of Delta load to Wye connection}
\begin{figure}[t]
    \centering
    \includegraphics[width=0.9\linewidth]{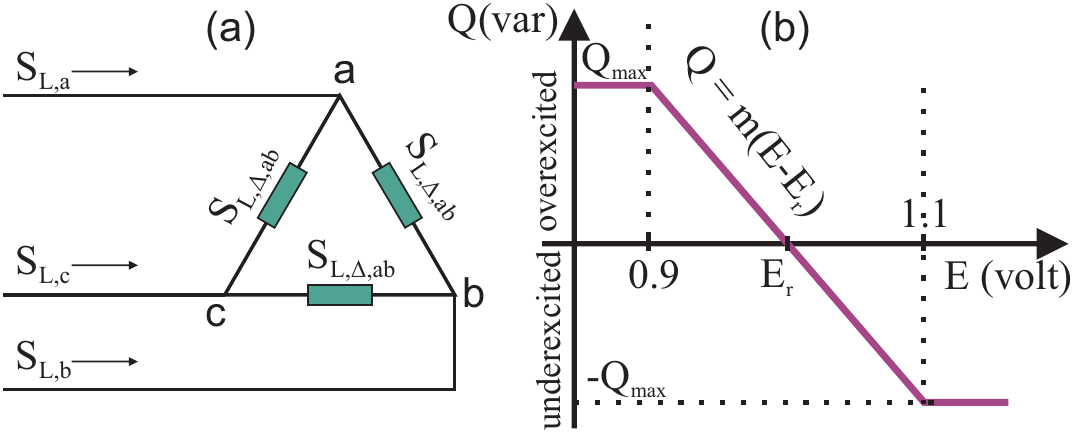}
       \caption{(a) 3-$\phi$ Delta load. (b) Volt-var characteristic of DER inverter.} \label{fig_volt_var}
\end{figure}

A Delta-connected load ($\mathbf{\bar{S}}_{L,\Delta}^i$) at bus $i$ can be transformed to equivalent Wye-connected ($\mathbf{\bar{S}}_{L,\Delta\text{-}\wye}^i$) load using transformation from \cite{Cheng2021AnNetworks}. With reference to Fig. \ref{fig_volt_var}(a), the transformation can be expressed as:
\begin{align}
\begin{bmatrix}
\bar{S}_{L,a}^i \\ \bar{S}_{L,b}^i \\ \bar{S}_{L,c}^i
\end{bmatrix} \hspace{-3pt}=\hspace{-3pt}
\begin{bmatrix}
\frac{\bar{E}_a^i}{\bar{E}_a^i-\bar{E}_b^i} & 0 & -\frac{\bar{E}_a^i}{\bar{E}_c^i-\bar{E}_a^i}\\
-\frac{\bar{E}_b^i}{\bar{E}_a^i-\bar{E}_b^i} & \frac{\bar{E}_b^i}{\bar{E}_b^i-\bar{E}_c^i} & 0\\
0 & -\frac{\bar{E}_c^i}{\bar{E}_b^i-\bar{E}_c^i} & \frac{\bar{E}_c^i}{\bar{E}_c^i-\bar{E}_a^i}
\end{bmatrix}&\hspace{-2pt}
\cdot \hspace{-2pt}
\begin{bmatrix}
\bar{S}_{L,\Delta,ab}^i\\ \bar{S}_{L,\Delta,bc}^i \\ \bar{S}_{L,\Delta,ca}^i
\end{bmatrix}\label{eq_delta_to_wye_load}\\
\text{In short form:}\quad \mathbf{\bar{S}}_{L,\Delta\text{-}\wye}^i =\bar{\Gamma}^i\cdot \mathbf{\bar{S}}_{L,\Delta}^i \quad \quad &
\end{align}
Note that (\ref{eq_delta_to_wye_load}) is also applicable for 1-$\phi$ or 2-$\phi$ Delta load by assuming them as 3-$\phi$ Delta load with 0 demand for the phases that are absent. A vector of Delta loads $\mathbf{\bar{S}}_{L,\Delta}=[\mathbf{\bar{S}}_{L,\Delta}^1, \ldots, \mathbf{\bar{S}}_{L,\Delta}^n]^T$ is converted to Wye-transformed load vector  $\mathbf{\bar{S}}_{L,\Delta\text{-}\wye}=[\mathbf{\bar{S}}_{L,\Delta\text{-}\wye}^1,\ldots,\mathbf{\bar{S}}_{L,\Delta\text{-}\wye}^n]$ as
\footnote{$Diag\{\}$ denotes a square diagonal matrix with the elements in $\{\}$ on the main diagonal.}:
\begin{align}
\mathbf{\bar{S}}_{L,\Delta\text{-}\wye} &= \bar{\mathbf{\Gamma}}\cdot \mathbf{\bar{S}}_{L,\Delta} \quad\text{where, } \bar{\mathbf{\Gamma}}=Diag\{\bar{\Gamma}^1,\ldots, \bar{\Gamma}^n\}. \label{eq_lod_delta_to_wye}
\end{align}
\subsubsection{Consideration of smart inverter functionality in DERs}
With the increasing adoption of the IEEE 1547-2018 standards by utilities, the DERs are required to provide voltage support to the grid by means of smart inverter functionality. The most commonly adopted functionality in DERs is volt-var support by which the DERs absorb or generate reactive power based on the voltage measured at its point of common coupling. An example of the volt-var characteristic of a smart inverter is shown in Fig. \ref{fig_volt_var}(b), by which the DER inverter provides dynamic reactive support based on the Q-V droop ($m$) when the terminal voltage is between 0.9 to 1.1 p.u.. With such smart inverter functionality, the imaginary component of complex power injection ($\bar{\mathbf{S}}_G$) in (\ref{eqn_Sinj}) depends on the magnitude of the terminal voltage. 
\par
(i) \textit{1-$\phi$ DERs:} Complex power injection of a single-phase DER connected to a node $(i,p)$ is expressed as\footnote{Imaginary unit of a complex number is denoted by $\bm{j}$ throughout the paper.}:
\begin{align}
    \bar{S}_{1\phi,p}^i = P_{1\phi,p}^i + \bm{j} m^i_{1\phi, p}(E_p^i-\hat{E}_{1\phi ,p}^i)
\end{align}
For generality, the 1-$\phi$ DER is assumed to exist on every node of the distribution network. In reality, DERs may not be at all nodes, however, the absence of a DER can be theoretically modeled with an inverter with 0 active power injection and 0 volt-var droop. Hence, the vector of complex power injection of all 1-$\phi$ inverters is expressed as:
\begin{align}
    \bar{\mathbf{S}}_{1\phi} = \mathbf{P}_{1\phi} + \bm{j}\mathbf{\Psi}\cdot(\mathbf{E}-\hat{\mathbf{E}}_{1\phi}) \label{eqn_sigle_ph_DER}
\end{align}
Here, $\mathbf{\Psi}=Diag\{m_{1\phi,a}^1,m_{1\phi,b}^1,m_{1\phi,c}^1,\ldots,m_{1\phi,a}^n,m_{1\phi,b}^n,\\m_{1\phi,c}^n\}$. Furthermore, $\mathbf{E}$  and $\hat{\mathbf{E}}$ are the vector of voltage magnitude and the reference value of voltages of all the nodes, respectively. For example, $\hat{\mathbf{E}}=[\hat{E}_{a}^1,\hat{E}_{b}^1,\hat{E}_{c}^1,\ldots,\hat{E}_{a}^n,\hat{E}_{b}^n,\hat{E}_{c}^n]^T$.

(ii) \textit{3-$\phi$ DERs:} For 3-$\phi$ inverter connected to bus $i$, the reactive power injection is computed from the similar volt-var characteristics utilizing the average voltage of all three nodes \cite{ElectricPowerResearchInstitute2013ModelingStudies}. Hence, the complex power injection of a three-phase inverter would be:
\begin{align}
     \bar{S}_{3\phi}^i = P_{3\phi}^i+\bm{j}m^i_{3\phi}\;[\frac{1}{3}(E_a^i+E_b^i+E_c^i)-\hat{E}_{3\phi}^i]
\end{align}
Consider every bus of the distribution network with a three-phase DERs with volt-var functionality, the vector of complex power injection from all the DERs would be
\begin{align}
    \bar{\mathbf{S}}_{3\phi} =\mathbf{P}_{3\phi}+\bm{j}(\mathbf{\Omega}\cdot\mathbf{E}-\mathbf{\Lambda}\cdot\hat{\mathbf{E}}_{3\phi}).\label{eq_S_3phaseDER}
\end{align}
The derivation of (\ref{eq_S_3phaseDER}) is detailed in Appendix \ref{app_3phaseDER}.
\par
Finally, using (\ref{eq_lod_delta_to_wye}), (\ref{eqn_sigle_ph_DER}) and (\ref{eq_S_3phaseDER}), the complex power injection model of the distribution network (\ref{eqn_Sinj}) can be formalized considering all bus modeled in the form of proposed composite form as:
\begin{align}
   & -\ubar{\mathbf{S}}_{L,\wye} - \ubar{\mathbf{\Gamma}}\cdot\ubar{\mathbf{S}}_{L,\Delta} + \ubar{\mathbf{S}}_{G}= \ubar{\mathbf{E}}\odot(\mathbf{\bar{Y}}^{o}+\mathbf{\bm{\delta}\bar{Y}^r})\cdot\mathbf{\bar{E}}\label{eqn_Sinj_2}
\end{align}
where,
\begin{align}
   &\ubar{\mathbf{S}}_{G} = \mathbf{P}_{1\phi}+ \mathbf{P}_{3\phi}- \bm{j}((\mathbf{\Psi}+\mathbf{\Omega})\cdot\mathbf{E}-(\mathbf{\Psi}\cdot\hat{\mathbf{E}}_{1\phi}+\mathbf{\Lambda}\cdot\hat{\mathbf{E}}_{3\phi})).\nonumber
\end{align}

\subsection{Derivation of sensitivity matrices}
In a distribution network, a node voltage depends on active/reactive power injections ($P^n_p/Q^n_p$) at any node $(n,p)$ and tap-position ($\gamma_s$) of voltage regulators. Hence, it is compelling to estimate the partial derivatives of voltage magnitude with respect to $u\in\{P^n_p, Q^n_p,\gamma_s\}$, where $(n,p)\in\mathcal{S}\cup\mathcal{C}$ and $s\in\mathcal{R}$. To find the general sensitivity equation, we assume a fictitious source at each node. Fictitious sources ($\bar{\mathbf{S}}_F$) do not exist in the network in reality (they can be visualized as a source injecting zero active/reactive power injection). However, they are used in the transmission system to study the impact on network voltage for a small change in power injection at a particular bus \cite{Machowski2008PowerDynamics}. The same concept of the fictitious source is utilized here and assumed to exist in each node of the distribution system to facilitate finding the derivative of node voltages with respect to active/reactive power injections. 
Hence, (\ref{eqn_Sinj_2}) is re-written as:
\begin{align}
     \ubar{\mathbf{S}}_{F}-\ubar{\mathbf{S}}_{L,\wye} - \ubar{\mathbf{\Gamma}}\cdot\ubar{\mathbf{S}}_{L,\Delta} + \ubar{\mathbf{S}}_{G} = \ubar{\mathbf{E}}\odot(\mathbf{\bar{Y}}^{o}+\mathbf{\bm{\delta}\bar{Y}^r})\cdot\mathbf{\bar{E}} \label{eqn_Sinj_3}.
\end{align}
For constant load models ($\mathbf{\bar{S}}_{L,\wye}$ and $\mathbf{\bar{S}}_{L,\Delta}$), the derivative of (\ref{eqn_Sinj_3}) with respect to $u$ can be expressed as:
\begin{align}
    &\frac{\partial \ubar{\mathbf{S}}_{F}}{\partial u}-\mathbf{\Pi}\cdot\frac{\partial \ubar{\mathbf{E}}}{\partial u} - \bm{j}(\mathbf{\Psi}+\mathbf{\Omega})\cdot\frac{\partial \mathbf{E}}{\partial u} \hspace{-3pt}=\hspace{-3pt} \frac{\partial \ubar{\mathbf{E}}}{\partial u} \odot  (\bar{\mathbf{Y}}^{o}\hspace{-2pt}+\hspace{-2pt}\bm{\delta}\bar{\mathbf{Y}}^r)\cdot\bar{\mathbf{E}} +\nonumber\\ &\ubar{\mathbf{E}}\odot (\bar{\mathbf{Y}}^{o}+\bm{\delta}\bar{\mathbf{Y}}^r)\cdot\frac{\partial \bar{\mathbf{E}}}{\partial u} +  
     \ubar{\mathbf{E}}\odot \frac{\partial \bm{\delta}\bar{\mathbf{Y}}^r}{\partial u}\cdot\bar{\mathbf{E}} \label{eq_sens_eqn}
\end{align}
Here, $\frac{\partial \ubar{\mathbf{\scriptstyle S}}_{L,\wye}}{\partial u} = 0$, $\frac{\partial}{\partial u}(\mathbf{\Gamma}\cdot\ubar{\mathbf{S}}_{\Delta}) =  \mathbf{\Pi}\cdot\frac{\partial \ubar{\mathbf{\scriptstyle E}}}{\partial u}$ and  $\frac{\partial \ubar{\mathbf{\scriptstyle S}}_{G}}{\partial u} = - \bm{j}(\mathbf{\Psi}+\mathbf{\Omega})\cdot\frac{\partial \mathbf{E}}{\partial u}$. The former expression is true for constant load models whereas the proof of the second expression is shown in Appendix \ref{app_wye_trans_del_load}. It is to be noted that $\frac{\partial}{\partial u}(\mathbf{\Gamma}\cdot\ubar{\mathbf{S}}_{\Delta})$ depends on derivative of voltage phasors whereas  $\frac{\partial \ubar{\mathbf{\scriptstyle S}}_{G}}{\partial u}$ depends on voltage magnitude. Hence, we express voltage phasor in terms of magnitude and angle. The voltage vector and its conjugate can be expressed as $\bar{\mathbf{E}}=\mathbf{E}\odot \bar{\mathbf{A}}$ and $\ubar{\mathbf{E}}=\mathbf{E}\odot \ubar{\mathbf{A}}$, where $\bar{\mathbf{A}}=[e^{\bm{j}\theta_a^1},e^{\bm{j}\theta_b^1},e^{\bm{j}\theta_c^1},\dots,e^{\bm{j}\theta_a^n},e^{\bm{j}\theta_b^n},e^{\bm{j}\theta_c^n}]^T$. Their derivatives with respect to $u$ can be further expanded as: 
\begin{subequations}\label{eq_dE_du}
\begin{align}
    \frac{\partial \bar{\mathbf{E}}}{\partial u}&=\frac{\partial \mathbf{E}}{\partial u}\odot \bar{\mathbf{A}}+\bm{j}\bar{\mathbf{E}}\odot\frac{\partial \boldsymbol{\theta}}{\partial u}\\
    \frac{\partial \ubar{\mathbf{E}}}{\partial u}&=\frac{\partial \mathbf{E}}{\partial u}\odot \ubar{\mathbf{A}}-\bm{j}\ubar{\mathbf{E}}\odot\frac{\partial \boldsymbol{\theta}}{\partial u}
\end{align}
\end{subequations}
It is to be noted that $\frac{\partial \bar{\mathbf{A}}}{\partial u}$ is substituted with $\frac{\partial \bar{\mathbf{A}}}{\partial u}=\bm{j}\bar{\mathbf{A}}\odot\frac{\partial\boldsymbol{\theta}}{\partial u}$. On substituting (\ref{eq_dE_du}) into (\ref{eq_sens_eqn}), we get:
\begin{align}
     \bar{\mathbf{F}} = & \bar{\mathbf{C}}\cdot\frac{\partial \mathbf{E}}{\partial u} + \bm{j}\bar{\mathbf{D}}\cdot\frac{\partial \boldsymbol{\theta}}{\partial u}.\label{eq_sf_sens_eqn}
\end{align}
where $\bar{\mathbf{C}} = \bar{\mathbf{C}}_1 + \bar{\mathbf{C}}_2 + \bar{\mathbf{C}}_3+\bar{\mathbf{C}}_4$ and $\bar{\mathbf{D}} = \bar{\mathbf{D}}_1 + \bar{\mathbf{D}}_2 + \bar{\mathbf{D}}_3$ such that:
\begin{align}
    \bar{\mathbf{C}}_1 =& Diag\{\ubar{\mathbf{A}} \odot (\bar{\mathbf{Y}}^{o}+\bm{\delta}\bar{\mathbf{Y}}^r)\cdot\bar{\mathbf{E}}\}
\end{align}
\begin{align}
    \bar{\mathbf{C}}_2 = & \ubar{\mathbf{E}}\odot (\bar{\mathbf{Y}}^{o}+\bm{\delta}\bar{\mathbf{Y}}^r)\cdot Diag\{\bar{\mathbf{A}}\}\\
    \bar{\mathbf{C}}_3 = & -\mathbf{\Pi}\cdot Diag\{\ubar{\mathbf{A}}\}, \quad \bar{\mathbf{C}}_4 = -\bm{j} (\mathbf{\Psi}+\mathbf{\Omega})\\
    \bar{\mathbf{D}}_1 =& -Diag\{\ubar{\mathbf{E}}\odot(\bar{\mathbf{Y}}^{o}+\bm{\delta}\bar{\mathbf{Y}}^r)\cdot\bar{\mathbf{E}}\}\\
    \bar{\mathbf{D}}_2 =& \ubar{\mathbf{E}}\odot (\bar{\mathbf{Y}}^{o}+\bm{\delta}\bar{\mathbf{Y}}^r)\cdot Diag\{\bar{\mathbf{E}}\}\\
     \bar{\mathbf{D}}_3 =&\mathbf{\Pi}\cdot Diag\{\ubar{\mathbf{E}}\}\\
    \bar{\mathbf{F}}=&  \frac{\partial \ubar{\mathbf{S}}_{F}}{\partial u} - \ubar{\mathbf{E}}\odot \frac{\partial \bm{\delta}\bar{\mathbf{Y}}^r}{\partial u}\cdot\bar{\mathbf{E}} \label{eqn_F}
\end{align}
The matrix $\frac{\partial \mathbf{E}}{\partial u}$ and $\frac{\partial \boldsymbol{\theta}}{\partial u}$ in (\ref{eq_sf_sens_eqn}) are real, and this equation can be segregated into two by equating real and imaginary components. Then the resulting equation in augmented matrix form is expressed as:
\begin{align}
    \begin{bmatrix}
    \Re(\bar{\mathbf{C}})&-\Im(\bar{\mathbf{D}})\\[1.2mm]
    \Im(\bar{\mathbf{C}})&\Re(\bar{\mathbf{D}})
    \end{bmatrix}
    \begin{bmatrix}
    \frac{\partial \mathbf{E}}{\partial u}\\[1.2mm]
    \frac{\partial \boldsymbol{\theta}}{\partial u}
    \end{bmatrix}
    =
    \begin{bmatrix}
    \Re(\bar{\mathbf{F}})\\[1.2mm]
    \Im(\bar{\mathbf{F}})
    \end{bmatrix}\label{eq_sens_matrix}
\end{align}
Finally (\ref{eq_sens_matrix}) can be solved for computation of voltage sensitivity with respect to any input variable $u$. It is worth noting that matrices $\bar{\mathbf{C}}$ and $\bar{\mathbf{D}}$ are constant matrices for a given operating condition, whereas only $\bar{\mathbf{F}}$ depends on the input variable $u$.

\subsubsection{Voltage sensitivity to tap-position of regulators ($u =\gamma$)}
The voltage sensitivity of all the nodes to a tap-position of any regulator can be determined considering $u=\gamma$ in (\ref{eq_sf_sens_eqn}) and its solution is given by (\ref{eq_sens_matrix}). For $u=\gamma$, $\frac{\partial \ubar{\mathbf{\scriptstyle S}}_{F}}{\partial \gamma}=0$ and $\bar{\mathbf{F}}$ in (\ref{eqn_F}) is simplified as:
\begin{align}
    \bar{\mathbf{F}}=& - \ubar{\mathbf{E}}\odot \frac{\partial \bm{\delta}\bar{\mathbf{Y}}^r}{\partial \gamma}\cdot\bar{\mathbf{E}} \label{eq_F_for_reg}
\end{align}
Here $\bar{\mathbf{F}}$ depends on the admittance matrix of regulators in the distribution network. $\bar{Y}^r$, ${\delta}\bar{{Y}}^r$, and $\frac{\partial {\delta}\bar{{Y}}^r}{\partial \gamma}$ for few type of regulators are shown in Appendix \ref{app_del_Y_1_phase}, \ref{app_del_Y_3_phase}. Thereafter, the voltage sensitivity matrix to a particular regulator can be determined by solving (\ref{eq_sens_matrix}). If there are $s$ regulators, one way of estimating the voltage sensitivity of the network is by solving a new set of (\ref{eq_sens_matrix}) for each regulator's tap-position.  For each regulator, the square matrix on the left side of (\ref{eq_sens_matrix}) is fixed whereas the right side matrix has to be computed using (\ref{eq_F_for_reg}). Alternatively, the voltage sensitivity with respect to all the regulators can be determined at once using the augmenting form of (\ref{eq_sens_matrix}), as shown below.
\begin{align}
    \begin{bmatrix}
    \Re(\bar{\mathbf{C}})\hskip-\arraycolsep&\hskip-\arraycolsep-\Im(\bar{\mathbf{D}})\\[1.2mm]
    \Im(\bar{\mathbf{C}})\hskip-\arraycolsep&\hskip-\arraycolsep\Re(\bar{\mathbf{D}})
    \end{bmatrix}\hspace{-4pt}
    \begin{bNiceMatrix}
    \frac{\partial \mathbf{E}}{\partial \gamma_1}\hskip-\arraycolsep&\hskip-\arraycolsep\dots\hskip-\arraycolsep&\hskip-\arraycolsep\frac{\partial \mathbf{E}}{\partial \gamma_s}\\[1.2mm]
    \frac{\partial \boldsymbol{\theta}}{\partial \gamma_1}\hskip-\arraycolsep&\hskip-\arraycolsep\dots\hskip-\arraycolsep&\hskip-\arraycolsep\frac{\partial \boldsymbol{\theta}}{\partial \gamma_s}
    \end{bNiceMatrix} \hspace{-4pt}=\hspace{-4pt}
    \begin{bmatrix}
    \Re(\bar{\mathbf{F}}_1)\hskip-\arraycolsep&\hskip-\arraycolsep\dots\hskip-\arraycolsep&\hskip-\arraycolsep\Re(\mathbf{F}_s)\\[1.2mm]
    \Im(\bar{\mathbf{F}}_1) \hskip-\arraycolsep&\hskip-\arraycolsep \dots\hskip-\arraycolsep&\hskip-\arraycolsep\Im(\mathbf{F}_s)
    \end{bmatrix}\label{eq_main_sens_eqn_tap}
\end{align}

\subsubsection{Voltage sensitivity to active/reactive power injections ($u=P^n_p \text{ or } Q^n_p$)}
Voltage sensitivity to active power injection ($u=P^n_p$) from a particular  node $(n,p)$ is also determined using (\ref{eq_sens_matrix}). For $u=P^n_p$, $\bar{\mathbf{F}}$ can be computed using (\ref{eqn_F}) and would be $\bar{\mathbf{F}}=\frac{\partial \ubar{\mathbf{\scriptstyle S}}_{F}}{\partial P^n_p}$ as $\frac{\partial \Delta\bar{\mathbf{Y}}^r}{\partial P^n_p}=\mathbf{0}$. Furthermore, $\bar{\mathbf{F}}$ can be deduced as:
\begin{align}
    \bar{F}^i_k=\frac{\partial \ubar{{S}}_{F_k}^i}{\partial P^n_p}=
    \begin{cases}
    1, & \text{if } (i,k)=(n,p)\\
    0, & \text{otherwise}
    \end{cases}, \forall (i,k) \in \mathcal{C} \label{eq_F}
\end{align}
The voltage sensitivity to active power injection at each node $(n,p)\in\mathcal{C}$ can be found by solving (\ref{eq_sens_matrix}) using (\ref{eq_F}) one by one. It can be noted that for any node injections, the square matrix on the left side of (\ref{eq_sens_matrix}) remains constant. Hene, we express the voltage sensitivity to active power injection from all the nodes in $\mathcal{C}$ by an augmented form as:

\begin{align}
    \begin{bmatrix}
    \Re(\bar{\mathbf{C}})&-\Im(\bar{\mathbf{D}})\\[1.2mm]
    \Im(\bar{\mathbf{C}})&\Re(\bar{\mathbf{D}})
    \end{bmatrix}
    \begin{bmatrix}
    \frac{\partial \mathbf{E}}{\partial P^1_a}&\dots&\frac{\partial \mathbf{E}}{\partial P^n_c}\\[1.2mm]
    \frac{\partial \boldsymbol{\theta}}{\partial P^1_a}&\dots&\frac{\partial \mathbf{\theta}}{\partial P^n_c}
    \end{bmatrix} =
    \begin{bmatrix}
    \mathbb{1}_{N\times N}\\[1.2mm]
    \mathbb{0}_{N\times N}
    \end{bmatrix},\label{eq_main_sens_eqn_P}
\end{align}
where $\mathbb{1}$ and $\mathbb{0}$ are an identity and zero matrix respectively.
\par
Similarly, the sensitivity matrix to reactive power injection is expressed as:
\begin{align}
    \begin{bmatrix}
    \Re(\bar{\mathbf{C}})&-\Im(\bar{\mathbf{D}})\\[1.2mm]
    \Im(\bar{\mathbf{C}})&\Re(\bar{\mathbf{D}})
    \end{bmatrix}
    \begin{bmatrix}
    \frac{\partial \mathbf{E}}{\partial Q^1_a}&\dots&\frac{\partial \mathbf{E}}{\partial Q^n_c}\\[1.2mm]
    \frac{\partial \boldsymbol{\theta}}{\partial Q^1_a}&\dots&\frac{\partial \mathbf{\theta}}{\partial Q^n_c}
    \end{bmatrix} =
    \begin{bmatrix}
    \mathbb{0}_{N\times N}\\[1.2mm]
    -\mathbb{1}_{N\times N}
    \end{bmatrix}\label{eq_main_sens_eqn_Q}
\end{align}

\subsubsection{Consideration of Slack buses}
The derivation of network sensitivity matrix in (\ref{eq_main_sens_eqn_tap}),(\ref{eq_main_sens_eqn_P}), and (\ref{eq_main_sens_eqn_Q}) considered all the buses to be of composite nature. All these equations can be expressed in short form as:
\begin{align}
\begin{bmatrix}
\mathbf{A}_1 & \mathbf{A}_2
\end{bmatrix}
    \cdot 
    \begin{bmatrix}
    \mathbf{X}_1\\
    \mathbf{X}_2
    \end{bmatrix}=
    \begin{bmatrix}
    \mathbf{B}_1\\\mathbf{B}_2
    \end{bmatrix} \label{eq_sens_short_form}
\end{align}
where $\mathbf{X}_1$ and $\mathbf{X}_2$ represent the voltage and angle sensitivity matrices, whereas $\mathbf{A}_1,\,\mathbf{A}_2,\,\mathbf{B}_1$ and $\mathbf{B}_2$ are constant matrices. The distribution networks have at least one slack bus and it is important to consider the slack nodes before solving (\ref{eq_sens_short_form}). For any slack node $(i, p)$, we can infer the following conditions.
\begin{align}
    \frac{\partial E_p^i}{\partial u} = 0 \text{ and } \frac{\partial \theta_p^i}{\partial u} = 0 \quad \forall (i,p)\in \mathcal{S}\label{eq_cond1}
\end{align}
To incorporate the nature of slack nodes in (\ref{eq_sens_short_form}), we represent the slack nodes' ($\mathcal{S}$) indices in $\mathbf{E}$ vector by a set $\mathcal{I}$. To incorporate (\ref{eq_cond1})  in (\ref{eq_sens_short_form}), we impose the following conditions.
\begin{align}
\mathbf{A}_1(i,k) \text{ and } \mathbf{A}_2(i,k) =& 
\begin{cases}
 1, \quad i=k\\
 0, \quad \text{otherwise}
\end{cases}\; \forall i,k\in \mathcal{I}\\
\mathbf{B}_1(i,k) \text{ and } \mathbf{B}_2(i,k) =& 0\quad \forall i\in \mathcal{I}
\end{align}

\subsubsection{Consideration of PV bus/node}
Unlike transmission networks, PV buses (where the voltage is held constant to a fixed value) are comparatively rare in distribution networks. A bus connected with a very large DER may be operated as a PV bus in the distribution network, however, the DER would require a large reactive power capability. Nevertheless, the proposed composite bus can be modeled as a PV bus by considering the connection of zero loads and a DER with a very large (theoretically infinite) volt-var droop.

\section{Numerical Test Cases}
To validate the proposed algorithm, we will demonstrate the results with visual and numerical verification. As sensitivities are basically the first-order partial derivatives of a system state to any input, we will show that the estimated sensitivities are tangential to non-linear state v.s. input plots at various operating points. For example, the voltage sensitivity of node $(i,p)$ to active power injection at node $(j,k)$, i.e., $\frac{\partial E^i_p}{\partial P^j_k}$ should be tangent to $E^i_p$ v.s. $P^j_k$ curves. Furthermore, we will compare our results with a perturb-\&-observe method for estimating the errors and benchmark with the existing analytical method \cite{Christakou2013EfficientNetworks}.
\subsection{Case study 1: IEEE 13 bus distribution network}
\begin{figure}
    \centering
    \includegraphics[width=\linewidth]{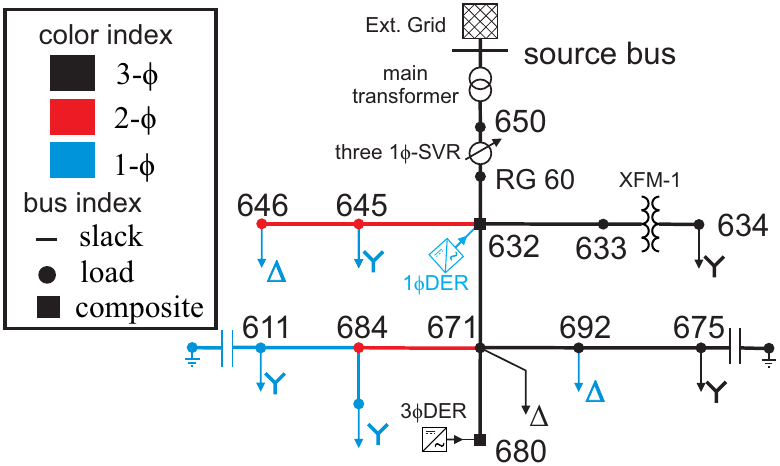}
    \vspace{-5mm}
       \caption{Modified IEEE 13 bus system} \label{fig_ieee13_bus}
\end{figure}
IEEE 13 bus network is an unbalanced distribution system with $1\text{-}\phi,2\text{-}\phi,$ and $3\text{-}\phi$ Delta and Wye connected loads as shown in Fig. \ref{fig_ieee13_bus}. It has only one substation transformer which is connected to a 115 kV transmission line at the source bus, which is the slack bus of the system, and has three $1\text{-}\phi$ SVRs at the substation. To test the proposed algorithm for estimating sensitivities, we created a composite bus at 632 and 680, where a 1-$\phi$ DER at node $a$ and a 3-$\phi$ DER both with volt-var control are connected, as shown in Fig. \ref{fig_ieee13_bus}.  

\subsubsection{Sensitivities to tap-position of regulators:}
\begin{figure}[t]
    \centering
    \includegraphics[width=\linewidth]{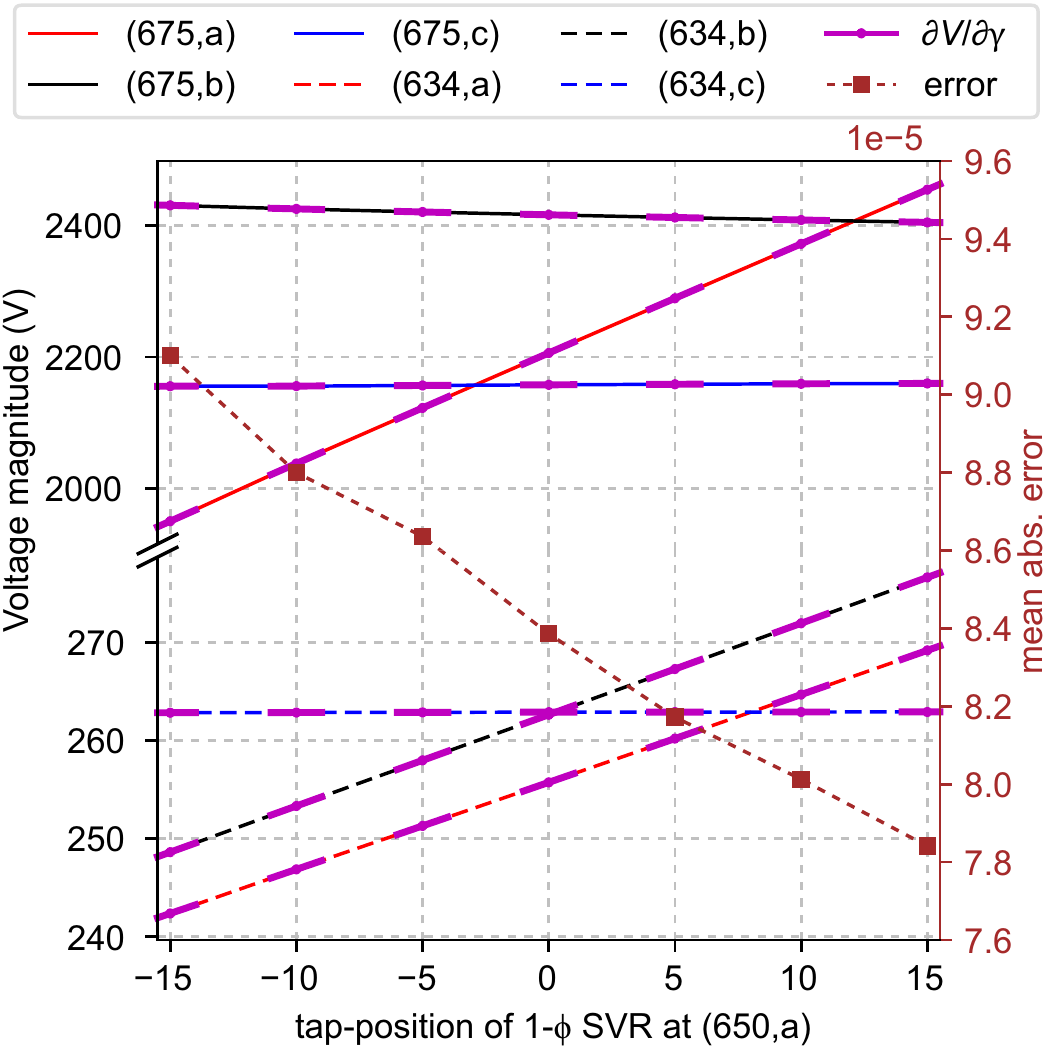}
    \vspace{-5mm}
       \caption{Plot showing the alignment of computed voltage sensitivities of buses 650 and 680 to tap-position of regulator1, with the corresponding V-tap curve in the IEEE 13 Bus network.}\label{fig_V_tap}
\end{figure}

\begin{figure}[t]
    \centering
    \includegraphics[width=\linewidth]{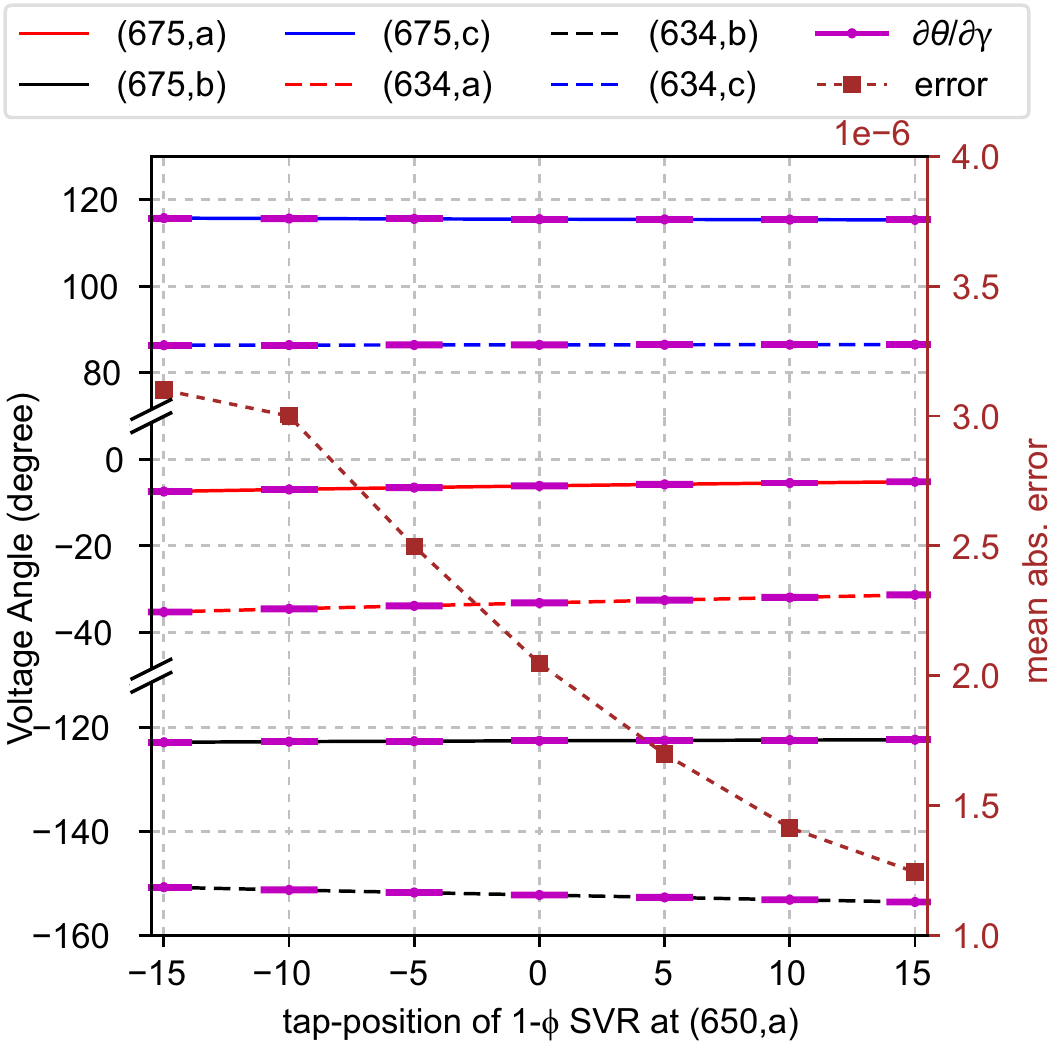}
    \vspace{-5mm}
       \caption{Plot showing the alignment of computed voltage sensitivities of buses 650 and 680 to tap-position of regulator1, with the corresponding $\theta$-tap curve in the IEEE 13 Bus network.}\label{fig_theta_tap}
\end{figure}

The proposed algorithm computed voltage and angle sensitivity matrix to tap-position of all three 1-$\phi$ SVR. However, we will first focus on sensitivity coefficients for the tap-position of SVR at phase $a$ for visual verification. To showcase visual verification, the voltage and angle at all nodes of bus 675 and 634 were recorded by solving distribution system power flow using OpenDSS at various tap-position of SVR located at (650,a). The recorded data have been presented in the form of solid lines for bus 675 and dashed lines for bus 634 in Fig. \ref{fig_V_tap} and Fig. \ref{fig_theta_tap}. Voltage and angle sensitivity coefficients to tap-position of SVR were extracted from the respective sensitivity matrices computed at various tap-position of SVR (e.g., -15,-10, -5, 0, 5, 10, 15). As the sensitivity coefficients are the slope of voltage and angle to tap-position, a small line with a corresponding slope at the operating point should be tangential to the plots obtained by a series of load flow computations from OpenDSS. Such small lines are shown in pink in Fig.  \ref{fig_V_tap} and Fig. \ref{fig_theta_tap}. These pink lines are visually tangential at every operating point under study, which supports the estimation accuracy of the proposed algorithm. 
\par
Furthermore, the computed voltage and angle sensitivity coefficients are compared with the perturb-\&-observe method. The mean absolute error for all the operating points shown in  Fig.  \ref{fig_V_tap} and Fig. \ref{fig_theta_tap} are below 9.2e-5 and 3.5e-6 for voltage and angle sensitivity estimation, respectively. From Fig. \ref{fig_V_tap}, one can see that when the tap-position of SVR is increased, it progressively increases the voltage at node (675,a), which is intuitive. However, it is not at all intuitive to observe that the voltage at both nodes (634,a) and (634,b) would increase with the tap-position of SVR. It is mainly because of the XFM-1 transformer with Delta-Wye configuration of HV and LV winding. It can also be noted from Fig. \ref{fig_theta_tap} that the tap-changes of SVR do not impact node angle significantly. 

\subsubsection{Sensitivities to active/reactive power injections:}
\begin{figure}[t]
    \centering
    \includegraphics[width=\linewidth]{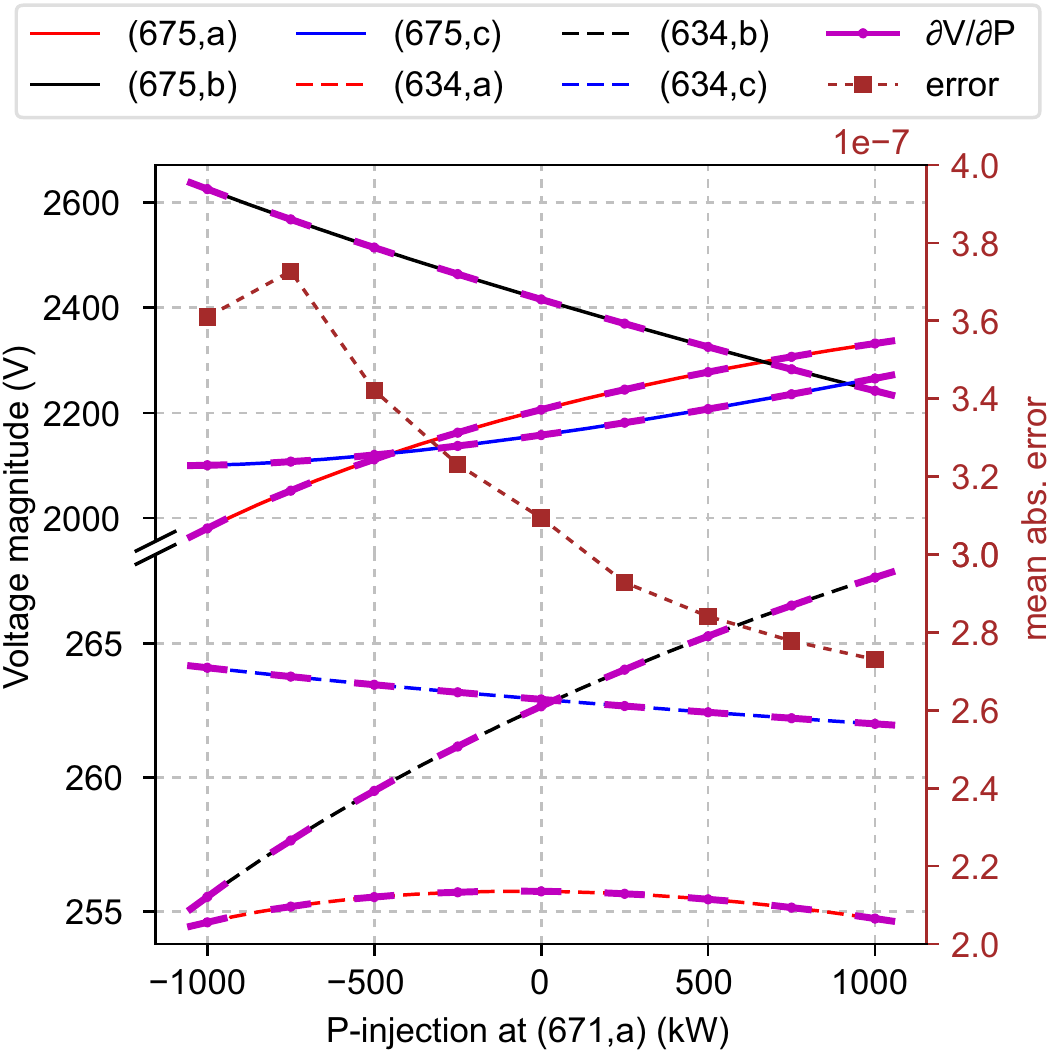}
    \vspace{-5mm}
    \caption{Plot showing the alignment of computed voltage magnitude sensitivities of buses 675 and 634 to active power injection at node (671,a), with the corresponding P-V curve in the IEEE 13 Bus network.}\label{fig_V_P}
\end{figure}

\begin{figure}[t]
    \centering
    \includegraphics[width=\linewidth]{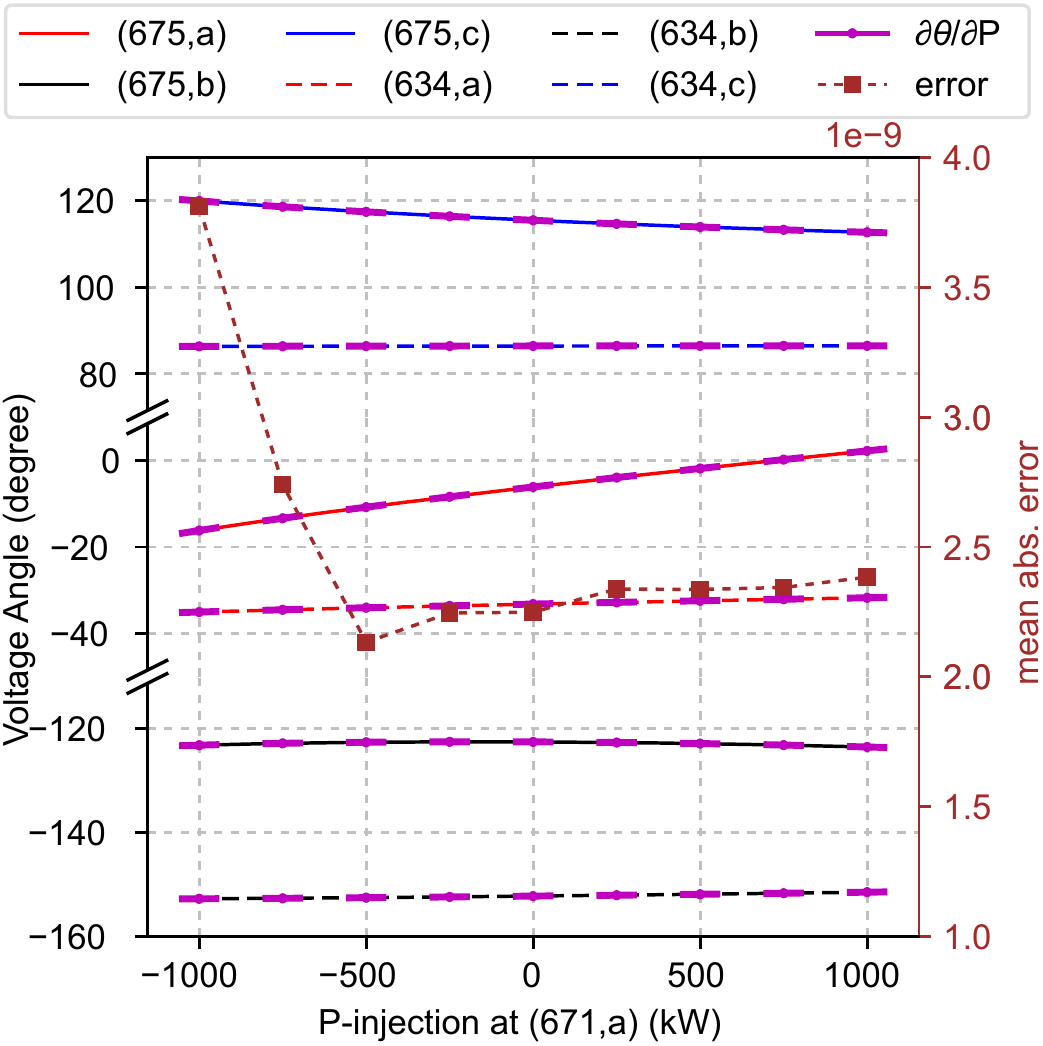}
    \vspace{-5mm}
    \caption{Plot showing the alignment of computed voltage angle sensitivities of buses 675 and 634 to active power injection at node (671,a), with the corresponding P-V curve in the IEEE 13 Bus network.}\label{fig_angle_P}
\end{figure}

\begin{figure}[t]
    \centering
    \includegraphics[width=\linewidth]{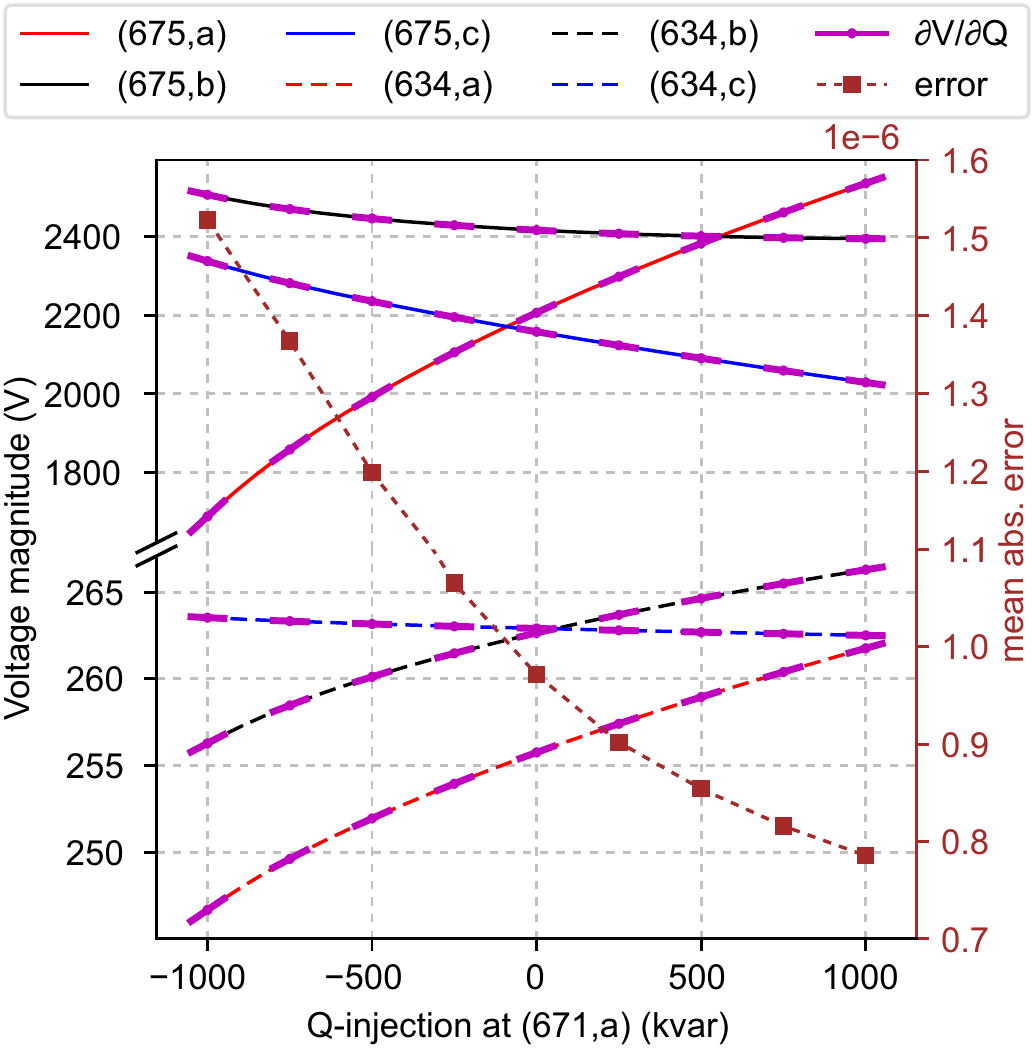}
    \vspace{-5mm}
    \caption{Plot showing the alignment of computed voltage magnitude sensitivities of buses 675 and 634 to reactive power injection at node (671,a), with the corresponding Q-V curve in the IEEE 13 Bus network.}\label{fig_V_Q}
\end{figure}

\begin{figure}[t]
    \centering
    \includegraphics[width=\linewidth]{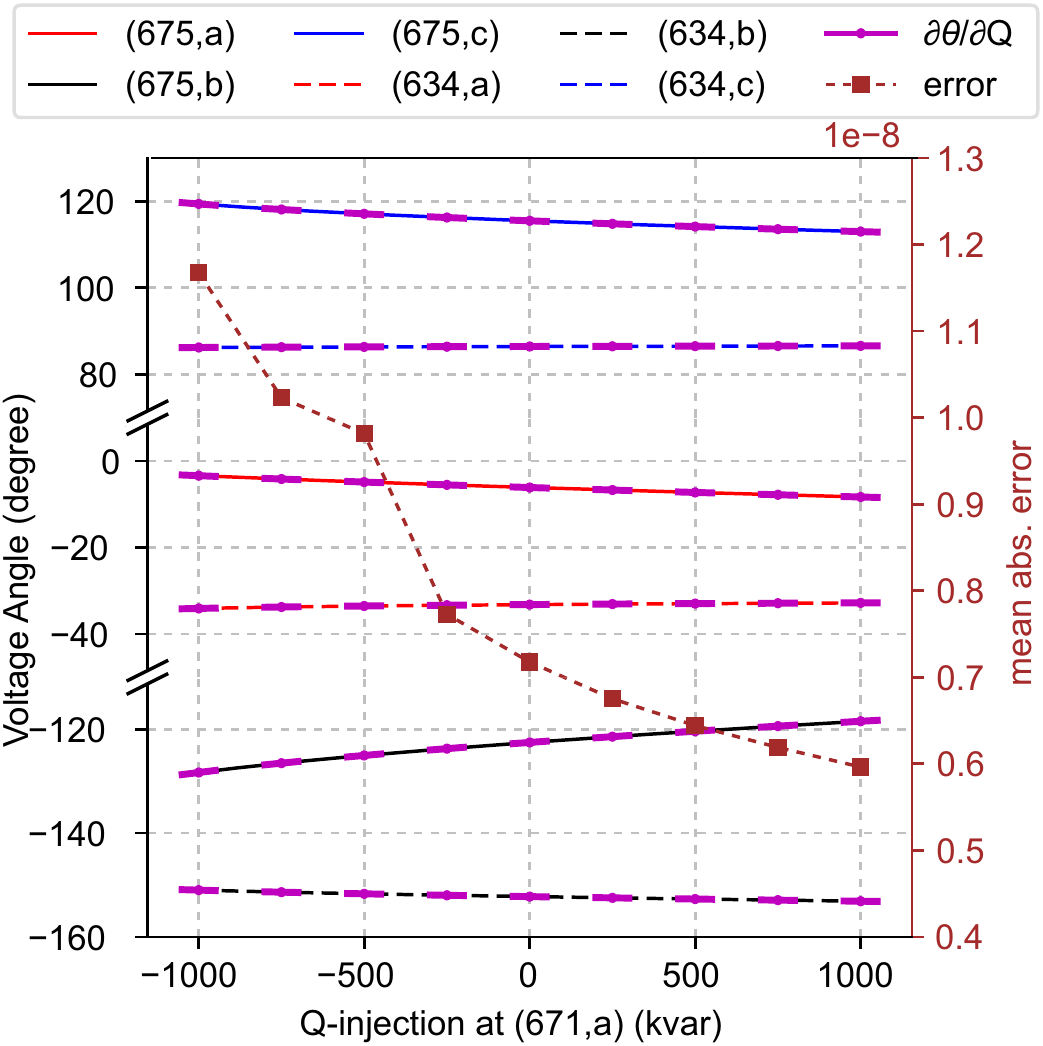}
    \vspace{-5mm}
    \caption{Plot showing the alignment of computed voltage angle sensitivities of buses 675 and 634 to reactive power injection at node (671,a), with the corresponding P-V curve in the IEEE 13 Bus network.}\label{fig_angle_Q}
\end{figure}

For visual verification, we will first focus on the voltage and angle trajectory of buses 675 and 634 to active/reactive power changes on the node (671,a). These trajectories were obtained by solving load flow problems in OpenDSS, which are shown using solid (for bus 675) and dashed lines (for bus 634) in Fig. \ref{fig_V_P}, Fig. \ref{fig_angle_P}, Fig.\ref{fig_V_Q} and Fig. \ref{fig_angle_Q}. It can be observed that the voltage and angle of nodes (675,a) and (634,a) changed significantly with active/reactive power injection in a node (671,a). However, other nodes at phase $b$ and $c$ showed slight changes. These plots show that a three-phased unbalanced distribution network is not intuitive because the phases are coupled by the mutual reactance of the lines. The proposed algorithm estimated the voltage and angle sensitivity matrix to active/reactive power injections at selected operating points, where the active/ reactive power injections at (671,a) are -1000, -750, -500,$\dots$, 1000 kW/kVar. From the computed sensitivity matrix, the sensitivity coefficients pertaining to power injection at (671,a) were located and used to draw a line having the slope equal to the sensitivity coefficients, and are shown by pink lines in Fig. \ref{fig_V_P} - \ref{fig_angle_Q}. All these lines are observed to be tangent to the corresponding plots obtained by a series of load flow calculations. 
\par
Alternatively, the computed sensitivity coefficients are compared with the perturb-\&-observed method, and the estimation errors were determined. The mean absolute error was determined at each operating point and is shown by the brown line in Fig. \ref{fig_V_P} - \ref{fig_angle_Q}. The estimated error is small and less than 1.6e-6 in all the operating points and all the plots.



\subsection{Case study 2: IEEE 123 bus distribution system}
\begin{figure}
    \centering
    \includegraphics[width=0.85\linewidth]{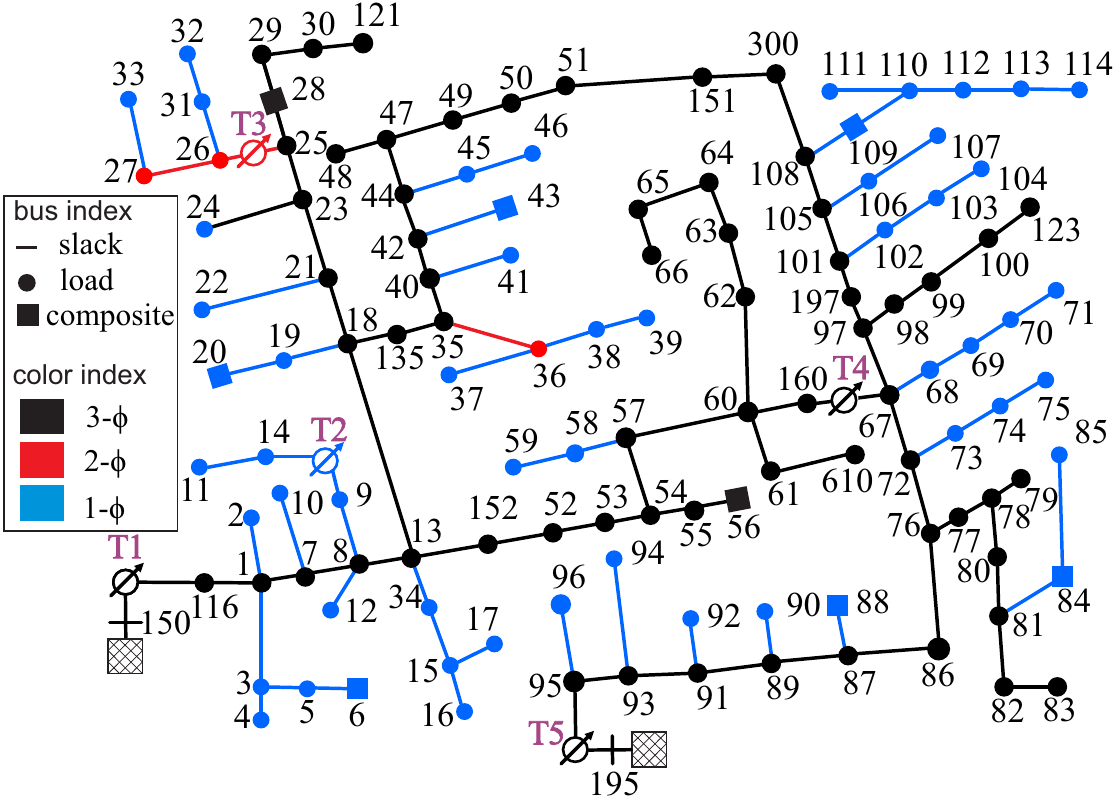}
    \vspace{-5mm}
       \caption{Modified IEEE 123 bus system} \label{fig_ieee123_bus}
\end{figure}

IEEE 123 bus distribution system is an unbalanced network with multiple substations, multiple line/substation regulators, 1-$\phi$, 2-$\phi$, and 3-$\phi$ Delta and Wye loads, as shown in Fig. \ref{fig_ieee123_bus}. In contrast to line regulators, the substation regulators are single 3-$\phi$ type, where the step change in tap-position changes all the phase voltages. To verify the capabilities of the proposed algorithm, the system is modified to have a ring configuration rather than a radial one by closing a switch between buses 151 and 300, as shown in Fig. \ref{fig_ieee123_bus}. In addition, six 1-$\phi$ DERs are connected at nodes (6,c), (88,a), (109,a), (84,c), (43,b), and (20,a), while two 3-$\phi$ DERs at buses 28 and 56. In Fig. \ref{fig_ieee123_bus}, the DER-connected buses are shown by a composite bus for clarity. 
\par
For this case study, the visual verification of computed sensitivity coefficients is illustrated only for the substation regulator at bus 150. This regulator is the single 3-$\phi$ type which was not present in IEEE 13 bus system studied above. Visual verification for sensitivity to active/reactive power changes is not presented deliberately because of space constraints. However, the numerical verification is studied in detail for all the cases in the subsequent subsection \ref{sec_perf_comp}.  
\par
\begin{figure}
    \centering
    \includegraphics[width=\linewidth]{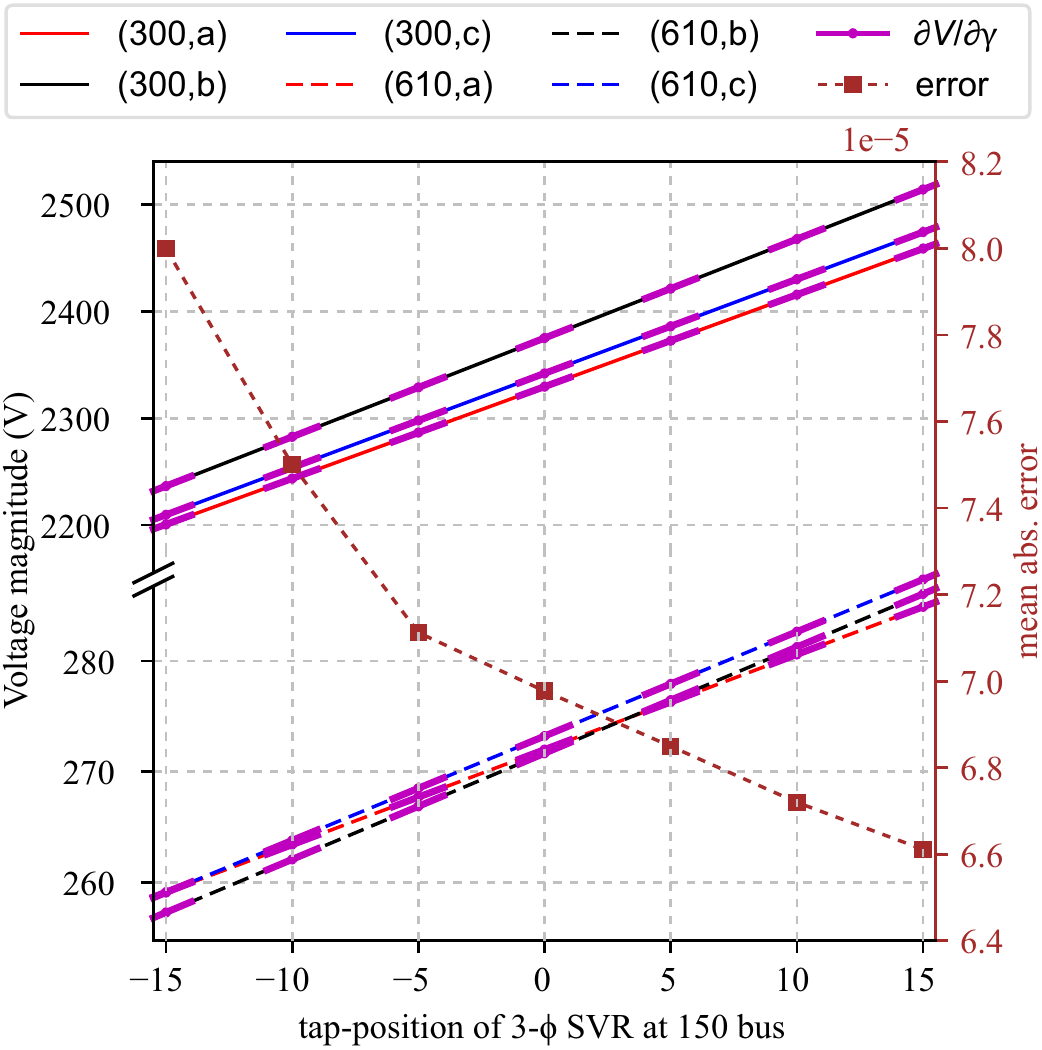}
    \vspace{-5mm}
       \caption{Plot showing the alignment of computed voltage sensitivities of buses 300 and 610 to tap-position of the 3-$\phi$ regulator at 150 bus, with the corresponding $\theta$-tap curve in the IEEE 123 Bus network.} \label{fig_V_tap_123_bus}
\end{figure}

The tap changes in 3-$\phi$ regulator change the voltage on all phases of the network. Assertively, Fig. \ref{fig_V_tap_123_bus} depicts the changes in voltage at buses 300 and 610 with a change in tap-position of 3-$\phi$ regulator located near bus 150. The sensitivity coefficients were determined by the proposed algorithm for the operating condition when the tap was at -15, -10,..., 10, 15. The obtained coefficients were used to draw a line at the corresponding operating point, and these lines are shown in pink color in the same Fig. \ref{fig_V_tap_123_bus}. All these lines are seen to be tangent, which provides visual confirmation of the accuracy of the proposed algorithm. Furthermore, the numerical verification of sensitivity coefficients was conducted by comparing with the perturb-\&-observe method, and the mean absolute errors were less than 8e-5 for all the cases shown in Fig. \ref{fig_V_tap_123_bus}.

\begin{table*}[t]
\small
  \centering
   \caption{Comparison of the proposed algorithm with analytical method}
    \begin{tabular}{|c|c|c|c|c|c|c|c|c|c|}
    \toprule
    \multicolumn{1}{|c|}{\multirow{4}[6]{1cm}{\centering case study}} & \multicolumn{1}{c|}{\multirow{4}[6]{1.4cm}{\centering sensitivity matrix}} & \multicolumn{4}{c|}{mean absolute percentage error} & \multicolumn{4}{c|}{mean computation time} \\
\cmidrule{3-10}          &       & \multicolumn{2}{c|}{IEEE 13 Bus} & \multicolumn{2}{c|}{IEEE 123 Bus} & \multicolumn{2}{c|}{IEEE 13 Bus} & \multicolumn{2}{c|}{IEEE 123 Bus} \\
\cmidrule{3-10}          &       & \multicolumn{1}{c|}{\multirow{2}[2]{1.4cm}{\centering Proposed method}} & \multicolumn{1}{c|}{\multirow{2}[2]{1.4cm}{\centering Analytical method}} & \multicolumn{1}{c|}{\multirow{2}[2]{1.4cm}{\centering Proposed method}} & \multicolumn{1}{c|}{\multirow{2}[2]{1.4cm}{\centering Analytical method}} & \multicolumn{1}{c|}{\multirow{2}[2]{1.4cm}{\centering Proposed method}} & \multicolumn{1}{c|}{\multirow{2}[2]{1.4cm}{\centering Analytical method}} & \multicolumn{1}{c|}{\multirow{2}[2]{1.4cm}{\centering Proposed method}} & \multicolumn{1}{c|}{\multirow{2}[2]{1.4cm}{\centering Analytical method}} \\
          &       &       &       &       &       &       &       &       &  \\
    \midrule
    \multicolumn{1}{|c|}{\multirow{6}[2]{*}{CS-A}} & $\partial\mathbf{V}/\partial\mathbf{P}$ & 0.02\% & 0.02\% & 0.04\% & 0.04\% & \multirow{6}[2]{*}{16 ms} & \multirow{6}[2]{*}{6.5 ms} & \multirow{6}[2]{*}{179.8 ms} & \multirow{6}[2]{*}{45.7ms} \\
          & $\partial\boldsymbol{\theta}/\partial\mathbf{P}$ & 0.05\% & N/A   & 0.07\% & N/A   &       &       &       &  \\
          & $\partial\mathbf{V}/\partial\mathbf{Q}$ & 0.09\% & 0.09\% & 0.10\% & 0.10\% &       &       &       &  \\
          & $\partial\boldsymbol{\theta}/\partial\mathbf{Q}$ & 0.01\% & N/A   & 0.02\% & N/A   &       &       &       &  \\
          & $\partial\mathbf{V}/\partial\boldsymbol{\gamma}$ & 0.01\% & N/A   & 0.06\% & N/A   &       &       &       &  \\
          & $\partial\boldsymbol{\theta}/\partial\boldsymbol{\gamma}$ & 0.01\% & N/A   & 0.05\% & N/A   &       &       &       &  \\
    \midrule
    \multicolumn{1}{|c|}{\multirow{6}[2]{*}{CS-B}} & $\partial\mathbf{V}/\partial\mathbf{P}$ & 0.02\% & 19.86\% & 0.11\% & 14.64\% & \multirow{6}[2]{*}{15.9 ms} & \multirow{6}[2]{*}{6.5 ms} & \multirow{6}[2]{*}{180.9 ms} & \multirow{6}[2]{*}{45.5 ms} \\
          & $\partial\boldsymbol{\theta}/\partial\mathbf{P}$ & 0.04\% & N/A   & 0.06\% & N/A   &       &       &       &  \\
          & $\partial\mathbf{V}/\partial\mathbf{Q}$ & 0.03\% & 84.11\% & 0.19\% & 52.91\% &       &       &       &  \\
          & $\partial\boldsymbol{\theta}/\partial\mathbf{Q}$ & 0.01\% & N/A   & 0.05\% & N/A   &       &       &       &  \\
          & $\partial\mathbf{V}/\partial\boldsymbol{\gamma}$ & 0.01\% & N/A   & 0.48\% & N/A   &       &       &       &  \\
          & $\partial\boldsymbol{\theta}/\partial\boldsymbol{\gamma}$ & 0.01\% & N/A   & 0.08\% & N/A   &       &       &       &  \\
    \midrule
    \multicolumn{1}{|c|}{\multirow{6}[2]{*}{CS-C}} & $\partial\mathbf{V}/\partial\mathbf{P}$ & 0.02\% & 21.32\% & 0.29\% & 18.23\% & \multirow{6}[2]{*}{16.1 ms} & \multirow{6}[2]{*}{6.5 ms} & \multirow{6}[2]{*}{180.5 ms} & \multirow{6}[2]{*}{45.5 ms} \\
          & $\partial\boldsymbol{\theta}/\partial\mathbf{P}$ & 0.04\% & N/A   & 0.35\% & N/A   &       &       &       &  \\
          & $\partial\mathbf{V}/\partial\mathbf{Q}$ & 0.08\% & 59.80\% & 0.51\% & 43.29\% &       &       &       &  \\
          & $\partial\boldsymbol{\theta}/\partial\mathbf{Q}$ & 0.02\% & N/A   & 0.18\% & N/A   &       &       &       &  \\
          & $\partial\mathbf{V}/\partial\boldsymbol{\gamma}$ & 0.01\% & N/A   & 0.67\% & N/A   &       &       &       &  \\
          & $\partial\boldsymbol{\theta}/\partial\boldsymbol{\gamma}$ & 0.01\% & N/A   & 0.10\% & N/A   &       &       &       &  \\
    \bottomrule
    \end{tabular}%
  \label{tab_comp}%
\end{table*}%

\subsection{Performance comparison and evaluations}\label{sec_perf_comp}
The previous subsection illustrated the performance of the proposed method pertaining to a few coefficients of sensitivity matrices. This subsection evaluates the estimated sensitivity matrices by determining their mean absolute percentage error (MAPE) with reference to sensitivity matrices computed using the perturb-and-observe method and mean computation time (MCT). To showcase our contribution, another analytical method \cite{Christakou2013EfficientNetworks} is also evaluated with a MAPE and MCT, and our performance is compared with it. Several case studies are studied in two test unbalanced distribution networks, such as IEEE 13 bus and IEEE 123 bus. Following are the details of case studies conducted on these networks.

\begin{enumerate}
    \item CS-A:  All loads are considered to be Wye-connected and DERs operated at a constant power factor.
    \item CS-B: Loads are either Delta- or Wye-connected and DERs operated at a constant power factor.
    \item CS-C: Loads are either Delta- or Wye-connected and DERs operated with volt-var control.
\end{enumerate}
Table \ref{tab_comp} shows the summary of the comparative study of the proposed method in terms of MAPE and MCT at different test distribution networks. For CS-A, all the sensitivity matrices estimated by the proposed method are almost the same as those estimated by an analytical method, however, the analytical method was not able to estimate a few sensitivity matrices such as $\partial\boldsymbol{\theta}/\partial\mathbf{P},\partial\boldsymbol{\theta}/\partial\mathbf{Q},\partial\mathbf{V}/\partial\boldsymbol{\gamma},$ and $\partial\boldsymbol{\theta}/\partial\boldsymbol{\gamma}$. In CS-B, where the Delta-connected loads are present, the analytical method show degraded performance with the MAPE of 19.86\% and 84.11\%  for the estimation of $\partial\mathbf{V}/\partial\mathbf{P}$ and $\partial\mathbf{V}/\partial\mathbf{Q}$, respectively. Whereas the proposed method performed equally well for CS-B as in CS-A. In CS-C, the Delta- and Wye-connected loads are the same as in CS-B, however, the DERs are operated with volt-var control. In this case study, the analytical method had degraded performance and the MAPE of 21.47\% and 43.35\% were seen for estimation of $\partial\mathbf{V}/\partial\mathbf{P}$ and $\partial\mathbf{V}/\partial\mathbf{Q}$ matrices, respectively. In contrast, the proposed method was better in the estimation of the sensitivity matrices albeit a small increment in MAPE was observed in comparison to CS-A and CS-B. The proposed estimation method performed  consistently better for all case studies CS-A, CS-B, and CS-C even for a larger test case, IEEE 123 Bus. The performance comparison in terms of MAPE and MCT for IEEE 123 Bus with the analytical method is highlighted in Table \ref{tab_comp}.

\section{Conclusion}
This paper proposed an analytical matrix-based method for the estimation of voltage magnitude and angle sensitivities to active/reactive and tap-position of the step voltage regulator (SVR) in unbalanced distribution networks. The proposed method is capable of estimating the sensitivity matrices to the tap-position of all line/substation regulators. Additionally, it is applicable for an unbalanced network with both Delta- and Wye-connected loads and with DERs having smart inverter functionality such as volt-var or power factor control. The reason behind such generic applicability of the proposed method is due to the composite bus modeling of each bus, which can be further deduced or simplified to any specific case of $1-\phi$, $2-\phi$, or $3-\phi$ Delta/Wye loads/DERs and their combination. 
\par
The proposed method is tested in unbalanced distribution test networks such as IEEE 13 bus and IEEE 123 bus which consider radial and ring topology (multiple slack buses), respectively. The accuracy of the proposed method is evaluated by computing a mean absolute percentage error with reference to the perturb-\&-observe method and by mean computation time. Additionally, the proposed method is compared with another analytical method for benchmarking. Compared to the other analytical method, the proposed method is more accurate in the presence of Delta-connected loads and DERs with volt-var control. The mean error of the proposed method is less than 0.7\% for all the case studies and for both the test distribution networks. 

\par

\appendix
In all the matrices listed below, $\bar{y}_T$ is short circuit impedance, $t_1/t_2$ is the tap ratio, $\gamma$ is a tap number, and $\Delta_K$ is the step voltage change of the regulator. 
\subsection{$\bar{Y}^r$, $\bar{\delta Y}^r$, and $\frac{\partial  \bar{\delta Y}^r}{\partial \gamma}$ for $1\text{-}\phi$ SVR} \label{app_del_Y_1_phase}
\begin{figure}[t]
    \centering
    \includegraphics[width=0.85\linewidth]{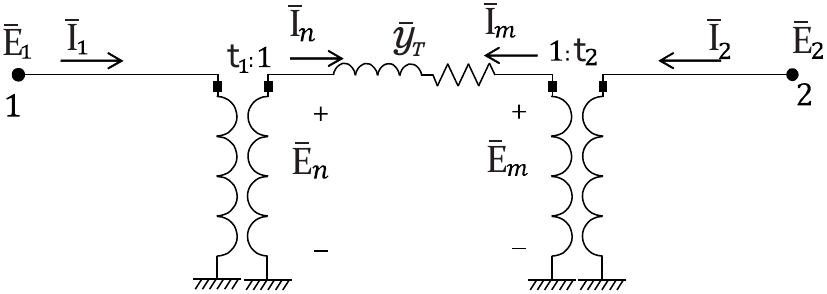}
    \vspace{-2mm}
    \caption{Two port model of SVR.}
    \label{fig_SVR}
\end{figure}

We consider the generic model of SVR that corroborates with OpenDSS, which comprises of tap setting at both 'from' and 'to' sides, as shown in Fig. \ref{fig_SVR}. The admittance matrix of SVR is expressed as:
\begin{align}
\bar{Y}^r &= 
\begin{bmatrix}
\frac{\bar{y}_T}{t_1^2} & -\frac{\bar{y}_T}{t_1t_2}\\
-\frac{\bar{y}_T}{t_1t_2} & \frac{\bar{y}_T}{t_2^2}
\end{bmatrix} \label{eq_SVR_Ymat}
\end{align}
\subsubsection{$1\text{-}\phi$ SVR with tap setting at 'from' side}
To model an SVR with tap provision at the 'from' side, we set $t_1=1+\gamma\Delta_K $ and $t_2=1$. With this we can obtain its $\bar{Y}^r$ from (\ref{eq_SVR_Ymat}) and its $\bar{\delta Y}^r$, and $\frac{\partial  \bar{\delta Y}^r}{\partial \gamma}$ are expressed as:
\begin{align}
\delta\bar{Y}^r &= \bar{y}_T
\begin{bmatrix}
\frac{1}{t_1^2}-1 & 1-\frac{1}{t_1}\\
1-\frac{1}{t_1} & 0
\end{bmatrix},\:
\frac{\partial \delta \bar{Y}^r}{\partial \gamma} = \bar{y}_T\Delta_K
\begin{bmatrix}
-\frac{2}{t_1^3} & \frac{1}{t_1^2}\\
\frac{1}{t_1^2} & 0
\end{bmatrix}
\end{align}

\subsubsection{$1\text{-}\phi$ SVR with tap setting at 'to' side}
Here, we set $t_1=1$ and $t_2=1+\gamma \Delta_k$, to obtain its $\bar{\delta Y}^r$, and $\frac{\partial  \bar{\delta Y}^r}{\partial \gamma}$ as:
\begin{align}
\delta\bar{Y}^r &= \bar{y}_T
\begin{bmatrix}
0 & 1-\frac{1}{t_2}\\
1-\frac{1}{t_2} & \frac{1}{t_2^2}-1
\end{bmatrix}
,\:\frac{\partial \delta\bar{Y}^r}{\partial \gamma} = \bar{y}_T \Delta_K
\begin{bmatrix}
0 & \frac{1}{t_2^2}\\
\frac{1}{t_2^2} & -\frac{2}{t_2^3}
\end{bmatrix}
\end{align}

\subsection{$\bar{Y}^r$, $\bar{\delta Y}^r$, and $\frac{\partial  \bar{\delta Y}^r}{\partial \gamma}$ for Wye-Wye $3\text{-}\phi$ SVR} \label{app_del_Y_3_phase}
Again, we consider the generic model of $3\text{-}\phi$ SVR that corroborates with OpenDSS, which comprises of tap setting at both 'from' and 'to' sides. The key difference between three $1\text{-}\phi$ and $3\text{-}\phi$ SVR is that individual phase voltage could be controlled in the former one whereas all phases are affected when the tap-position is changed in the latter one. The admittance matrix of Wye-Wye connected $3\text{-}\phi$ SVR is expressed as:
\begin{align}
\bar{Y}^r \hspace{-1.2mm}= \hspace{-1.8mm}
\begin{bmatrix}
\frac{\bar{y}_T}{t_1^2} & 0 & 0 & -\frac{\bar{y}_T}{t_1t_2} & 0 & 0\\
0 & \frac{\bar{y}_T}{t_1^2} 0 & 0 & 0 & -\frac{\bar{y}_T}{t_1t_2} & 0\\
0 & 0 & \frac{\bar{y}_T}{t_1^2} &  0 & 0& -\frac{\bar{y}_T}{t_1t_2}\\
-\frac{\bar{y}_T}{t_1t_2} & 0 & 0 & \frac{\bar{y}_T}{t_2^2} & 0 & 0\\
 0 & -\frac{\bar{y}_T}{t_1t_2} & 0 &  0 & \frac{\bar{y}_T}{t_2^2} & 0\\
  0 &  0 & -\frac{\bar{y}_T}{t_1t_2} & 0 & 0 & \frac{\bar{y}_T}{t_2^2}\\
\end{bmatrix} \label{eq_3pSVR_Ymat}
\end{align}
$\bar{\delta Y}^r$ and $\frac{\partial  \bar{\delta Y}^r}{\partial \gamma}$ for Wye-Wye $3\text{-}\phi$ SVR can be obtained from (\ref{eq_3pSVR_Ymat}), following the steps shown for $1\text{-}\phi$ SVR in Appendix \ref{app_del_Y_1_phase}. 

\subsection{Complex power injection from 3-$\phi$ DERs}\label{app_3phaseDER}
The complex power injection of 3-$\phi$ inverter with volt-var functionality would be:
\begin{align}
     \bar{S}_{3\phi}^i = P_{3\phi}^i+\bm{j}m^i_{3\phi}\;[\frac{1}{3}(E_a^i+E_b^i+E_c^i)-\hat{E}_{3\phi}^i)]
\end{align}

Here,  $\bar{S}_{3\phi}^i$ is a total power that is divided uniformly among three phases by the inverter controllers \cite{ElectricPowerResearchInstitute2013ModelingStudies}. Hence, active and reactive power injection at each phase of bus $i$ would be:

\begin{align}
&\begin{bmatrix}
    P_{3\phi,a}^i,P_{3\phi,b}^i,P_{3\phi,c}^i
    \end{bmatrix}^T= \frac{1}{3}
    \begin{bmatrix}
    P_{3\phi}^i, P_{3\phi}^i, P_{3\phi}^i
    \end{bmatrix}^T
\end{align}
\begin{align}
     &\begin{bmatrix}
    Q_{3\phi,a}^i\\Q_{3\phi,b}^i\\Q_{3\phi,c}^i
    \end{bmatrix}= 
    \begin{bmatrix}
     \frac{m^i_{3\phi}}{3}(\frac{1}{3}(E_a^i+E_b^i+E_c^i)-\hat{E}_{3\phi}^i) \\ 
     \frac{m^i_{3\phi}}{3}(\frac{1}{3}(E_a^i+E_b^i+E_c^i)-\hat{E}_{3\phi}^i)\\
     \frac{m^i_{3\phi}}{3}(\frac{1}{3}(E_a^i+E_b^i+E_c^i)-\hat{E}_{3\phi}^i)
    \end{bmatrix}\\
    &=\hspace{-5pt}\begin{bmatrix}
    \frac{m^i_{3\phi}}{9} & \frac{m^i_{3\phi}}{9} & \frac{m^i_{3\phi}}{9}\\
    \frac{m^i_{3\phi}}{9} & \frac{m^i_{3\phi}}{9} & \frac{m^i_{3\phi}}{9}\\
    \frac{m^i_{3\phi}}{9} & \frac{m^i_{3\phi}}{9} & \frac{m^i_{3\phi}}{9}
    \end{bmatrix}\hspace{-2pt}\cdot\hspace{-2pt}
    \begin{bmatrix}
    E_a^i\\E_b^{i}\\E_c^{i}
    \end{bmatrix}\hspace{-3pt}-\hspace{-3pt}
    \begin{bmatrix}
    \frac{m^i_{3\phi}}{3} & 0 & 0\\
    0 &  \frac{m^i_{3\phi}}{3} & 0\\
    0 & 0 &  \frac{m^i_{3\phi}}{3}
    \end{bmatrix}\hspace{-2pt}\cdot\hspace{-2pt}
    \begin{bmatrix}
    \hat{E}_{3\phi}^i\\\hat{E}_{3\phi}^i\\\hat{E}_{3\phi}^i
    \end{bmatrix}\\
    &=\Omega^i\cdot\mathbf{E}^i-\Lambda^i\cdot\hat{\mathbf{E}}_{3\phi}^i
\end{align}
Hence, the vector of complex power injection from three-phase DERs at each bus would be:
\begin{align}
    \bar{\mathbf{S}}_{3\phi} =\mathbf{P}_{3\phi}+\bm{j}(\mathbf{\Omega}\cdot\mathbf{E}-\mathbf{\Lambda}\cdot\hat{\mathbf{E}}_{3\phi}).
\end{align}
where $\mathbf{\Omega}=Diag\{\Omega^1,\ldots,\Omega^n\}$ and $\mathbf{\Lambda}=Diag\{\Lambda^1,\ldots,\Lambda^n\}$
\subsection{Sensitivity of Wye-transformed Delta-connected loads}\label{app_wye_trans_del_load}
When constant power Delta-connected load at bus $i$ is transformed to Wye, the resulting Wye-connected loads ($\mathbf{\bar{S}}_{\Delta\text{-}\wye}^i$) become voltage dependent as shown by (\ref{eq_delta_to_wye_load}). The sensitivity of $\mathbf{\bar{S}}_{\Delta\text{-}\wye}^i$ with respect to any input variable $u$ is determine by differentiating (\ref{eq_delta_to_wye_load}), and on after rearranging, we get (\ref{eq_sens_of_delta_wye_load}). In short, (\ref{eq_sens_of_delta_wye_load}) can be expressed as:
\begin{align}
    \frac{\partial \mathbf{\bar{S}}_{\Delta\text{-}\wye}^i}{\partial u} = \bar{\Pi}^i \cdot \frac{\partial \mathbf{\bar{E}}^i}{\partial u}. \label{eq_sens_of_delta_wye_load_short_form}
\end{align}
Utilizing (\ref{eq_sens_of_delta_wye_load_short_form}), the sensitivity of the vector of Wye-transformed Delta loads can be expressed as:
\begin{align}
    \frac{\partial \mathbf{\bar{S}}_{\Delta\text{-}\wye}}{\partial u} = \bar{\mathbf{\Pi}} \cdot \frac{\partial \mathbf{\bar{E}}}{\partial u}.
\end{align}
where,
\begin{align}
    \mathbf{\bar{S}}_{\Delta\text{-}\wye} = [\mathbf{\bar{S}}_{\Delta\text{-}\wye}^1,\dots, \mathbf{\bar{S}}_{\Delta\text{-}\wye}^n]^T \quad \text{and}\quad \bar{\mathbf{\Pi}} = Diag\{\bar{\Pi}^1, \dots,\bar{\Pi}^n\}.
\end{align}

\begin{figure*}[b]
\centering
\small
\begin{align}
\begin{bmatrix}
\frac{\partial\bar{S}^i_{\Delta,a}}{\partial u} \\ \frac{\partial\bar{S}^i_{\Delta,b}}{\partial u} \\ \frac{\partial\bar{S}^i_{\Delta,c}}{\partial u}
\end{bmatrix} = -
\begin{bmatrix}
-\frac{\bar{S}^i_{ab}\bar{E}^i_b}{(\bar{E}^i_a-\bar{E}^i_b)^2} - \frac{\bar{S}^i_{ca}\bar{E}^i_c}{(\bar{E}^i_c-\bar{E}^i_a)^2}  & \frac{\bar{S}^i_{ab}\bar{E}^i_a}{(\bar{E}^i_a-\bar{E}^i_b)^2} & \frac{\bar{S}^i_{ca}\bar{E}^i_a}{(\bar{E}^i_c-\bar{E}^i_a)^2}\\
\frac{\bar{S}^i_{ab}\bar{E}^i_b}{(\bar{E}^i_a-\bar{E}^i_b)^2} & -\frac{\bar{S}^i_{ab}\bar{E}^i_a}{(\bar{E}^i_a-\bar{E}^i_b)^2} - \frac{\bar{S}^i_{bc}\bar{E}^i_c}{(\bar{E}^i_b-\bar{E}^i_c)^2}  & \frac{\bar{S}^i_{bc}\bar{E}^i_b}{(\bar{E}^i_b-\bar{E}^i_c)^2} \\
\frac{\bar{S}^i_{ca}\bar{E}^i_c}{(\bar{E}^i_c-\bar{E}^i_a)^2} & \frac{\bar{S}^i_{bc}\bar{E}^i_c}{(\bar{E}^i_b-\bar{E}^i_c)^2}    & -\frac{\bar{S}^i_{bc}\bar{E}^i_b}{(\bar{E}^i_b-\bar{E}^i_c)^2} - \frac{\bar{S}^i_{ca}\bar{E}^i_a}{(\bar{E}^i_c-\bar{E}^i_a)^2}
\end{bmatrix}
\cdot 
\begin{bmatrix}
\frac{\partial\bar{E}^i_{a}}{\partial u}\\ \frac{\partial\bar{E}^i_{b}}{\partial u} \\ \frac{\partial\bar{E}^i_{c}}{\partial u}
\end{bmatrix} \label{eq_sens_of_delta_wye_load}
\end{align} 
\end{figure*}

\ifCLASSOPTIONcaptionsoff
  \newpage
\fi
\bibliographystyle{IEEEtran}

\bibliography{bare_jrnl}

\end{document}